\documentclass[a4paper,fleqn]{cas-dc}

\usepackage{graphicx}                     
\usepackage[numbers,sort&compress]{natbib} 
\usepackage{subcaption}
\usepackage{amsmath}
\usepackage{booktabs}                     
\usepackage{xr-hyper}                     
\usepackage{hyperref}
\IfFileExists{supplementary.aux}{\externaldocument[sp-]{supplementary}}{}
\usepackage{float}
\usepackage{placeins}                    
\usepackage{enumitem}
\usepackage{xcolor}
\usepackage{amssymb}
\usepackage{geometry}
\usepackage{multirow}
\usepackage{array}

\begin{document}

\let\WriteBookmarks\relax
\def\floatpagepagefraction{1}
\def\textpagefraction{.001}

\ExplSyntaxOn
\cs_gset:Npn \__first_footerline:
  { \group_begin: \small \sffamily \__short_authors: \group_end: }
\bool_gset_true:N \g_stm_nologo_bool
\ExplSyntaxOff

\shorttitle{Chalcopyrite-inspired tellurides for solar cells}
\shortauthors{M.H.S. Rifat et al.}

\title [mode = title]{\centering Design of Chalcopyrite-Inspired Tellurides for Solar Cells:\\A First-Principles and Coupled FDTD--CHARGE Device Study}


\author[1]{Md Hasan Shahriar Rifat}
\orcidauthor{0009-0001-6032-5160}{Md Hasan Shahriar Rifat}
\ead{rifatmse@gmail.com}            
\credit{}
\cormark[1]
\affiliation[1]{{Materials Science and Engineering,},organization={University of Rajshahi},state={Rajshahi 6205},country={Bangladesh}}

\author[2]{Tanvir Khan}
\credit{}
\affiliation[2]{{Dept. of Physics,},organization={University of Rajshahi},state={Rajshahi 6205},country={Bangladesh}}

\author[3]{Ababil Hossain Fahad}
\credit{}
\affiliation[3]{{Department of Electrical and Electronic Engineering,},organization={BRAC University},state={Dhaka},country={Bangladesh}}

\author[1]{Md Arafat Hossain Shourov}
\credit{}

\author[1]{Md Sahat Bin Sayed}
\credit{}
\author[4]{K.M. Mehedi Hassan}
\credit{}
\affiliation[4]{{Department of Electrical and Electronic Engineering (EEE),},organization={Khulna University of Engineering and Technology},state={Khulna 9203},country={Bangladesh}}


\cortext[1]{Corresponding author}

\begin{abstract}
The defect-chalcopyrite tellurides XGa$_2$Te$_4$ (X = Cd, Zn) were evaluated as
photovoltaic absorbers by combining first-principles calculations with coupled optical
and electrical device simulations. Density functional theory with the GGA-PBEsol
functional yields tetragonal $I\bar{4}$ structures with negative formation energies and
phonon spectra free of imaginary modes. Both compounds have direct band gaps at $\Gamma$,
of 1.046 and 1.032~eV with PBEsol and 2.21 and 1.895~eV with the HSE06 hybrid functional,
and absorption coefficients exceeding $10^{5}$~cm$^{-1}$ in the visible range.
Wannier--Mott estimates give exciton binding energies of 18.85 and 26.81~meV, comparable
to $k_{\mathrm{B}}T$ at 300~K, and exciton Bohr radii of 34.3 and 23.0~\AA.
Constant-relaxation-time Boltzmann transport calculations give Seebeck coefficients of up
to $\approx$255~$\mu$V~K$^{-1}$ near 350--400~K. Mulliken-electronegativity estimates
place both band edges outside the water redox potentials at pH~0, with hydrogen-evolution
driving forces of 0.83 and 0.60~eV but oxygen-evolution margins of only 0.16 and
0.07~eV. Finite-difference time-domain (FDTD) optical simulations coupled to
drift--diffusion (CHARGE) electrical simulations of FTO/IGZO/XGa$_2$Te$_4$/CuAlO$_2$/Pt
cells reveal a small conduction-band spike at the front junction (+0.15 and +0.33~eV),
reflection losses governed by interference in the front oxide stack, electron collection
confined to $\approx$0.5~$\mu$m from the junction, and a fill-factor loss in the
CdGa$_2$Te$_4$ device caused by the CuAlO$_2$ hole contact. Because the imported optical
constants retain the PBEsol absorption edge, the simulated photocurrents exceed the
detailed-balance limit; absolute device efficiencies are therefore treated as upper
bounds, whereas the identified loss mechanisms and design trends are robust.
These results identify CdGa$_2$Te$_4$ and ZnGa$_2$Te$_4$ as candidate thin-film absorbers
and define the contact engineering required to exploit them.
\end{abstract}

\begin{keywords}
Defect chalcopyrite
\sep CdGa$_2$Te$_4$ and ZnGa$_2$Te$_4$
\sep Density functional theory
\sep Lumerical FDTD and CHARGE
\sep Optical and electronic properties
\sep Thermoelectric properties
\sep Photocatalysis
\end{keywords}

\sloppy
\maketitle

\section{Introduction}
\label{sec:intro}

Continued reliance on fossil fuels is incompatible with sustainable development, given the
rising global energy demand and the environmental impact of combustion, and solar energy
has therefore emerged as a leading clean and renewable alternative \cite{alshetwi2022}.
Over the past decade, perovskite solar cells (PSCs) have attracted particular attention
\cite{stranks2015,yang2015,ye2021,zhou2016,zhou2016a} because their
light-harvesting layers combine simple fabrication, a tunable band gap and long carrier
lifetimes \cite{li2017,min2021}, and their power conversion efficiency (PCE) has risen from
3.8\% to 25.73\% \cite{kojima2009,park2023}. These advances rest on a high absorption
coefficient, adjustable band gaps, long carrier diffusion lengths, low exciton binding
energies and high carrier mobilities \cite{stranks2015,li2017}. However, the lead (Pb)
contained in the most efficient compositions poses considerable environmental and health
risks \cite{babayigit2016,lyu2017,ren2022}, which motivates the search for efficient and
less hazardous absorbers \cite{khan2024}.

Density functional theory (DFT) has become a powerful tool for predicting the structural,
optical, mechanical and electronic properties of candidate photovoltaic (PV) materials
before synthesis \cite{pu2021,zahra2024,hossain2024}. Jahangirli \emph{et al.}, for
example, combined experiment and GGA-based DFT to study the electronic, optical and
vibrational properties of CdGa$_2$Te$_4$ and highlighted its potential for PV technologies
\cite{jahangirli2020}. CdGa$_2$Te$_4$ belongs to the defect chalcopyrites (DCs), a family
of compounds with the general formula AB$_2$X$_4$ (A = Zn, Cd or Hg; B = Ga or Al;
X = S, Se or Te) that are structurally derived from the chalcopyrites and crystallize in
a tetragonal structure under ambient conditions. DC semiconductors exhibit notable
electronic, elastic, optical and nonlinear properties, which make them candidates for
nonlinear optics, photovoltaic devices and photodetectors
\cite{tiginyanu2003,reshak2014,mayengbam2018}. Among them, CdGa$_2$Te$_4$ and
ZnGa$_2$Te$_4$ are prominent members with nearly ideal chalcopyrite-like structures.

The structural, electronic, optical and nonlinear properties of several DC compounds,
including CdGa$_2$X$_4$ (X = S, Se) \cite{ma2012,reshak2014}, XAl$_2$Se$_4$ (X = Cd,
Hg) \cite{verma2011,fuentescabrera2001,singh2013}, ZnAl$_2$Te$_4$ \cite{tripathy2018} and
ZnGa$_2$Te$_4$ \cite{xiaoshu2010,mayengbam2018}, have been examined
by first-principles calculations. Jiang and Lambrecht compared the band structures of DC
semiconductors with those of their parent chalcopyrites \cite{jiang2004}, and the
temperature and pressure dependence of the structural and electronic properties of
CdAl$_2$Se$_4$ has been computed \cite{singh2012}. Despite this body of work, many basic
properties of the tellurides remain insufficiently assessed, and their behavior inside a
complete solar cell has not been examined.

The performance of a thin-film solar cell depends on how the absorber is combined with its
charge-selective contacts. The electron-transport layer (ETL) and hole-transport layer
(HTL) must each extract one carrier type, block the other and form interfaces with low
recombination, while the front layers must transmit as much light as possible to the
absorber. The conduction-band offset at the absorber/ETL interface is particularly
important: a small spike suppresses interface recombination, whereas a cliff lowers the
open-circuit voltage ($V_{\mathrm{OC}}$) \cite{minemoto2015}. In this work, amorphous
indium gallium zinc oxide (IGZO) is used as the ETL because it is a wide-gap n-type oxide
with high electron mobility that can be deposited at room temperature \cite{nomura2004}.
Delafossite CuAlO$_2$, the first reported p-type transparent conducting oxide
\cite{kawazoe1997}, is used as the HTL, and fluorine-doped tin oxide (FTO) and Pt serve as
the front and back electrodes. Because every layer has a distinct optical or electrical
function, the role of the absorber can be isolated.

Here, we combine DFT with coupled optical and electrical device simulations to evaluate
CdGa$_2$Te$_4$ and ZnGa$_2$Te$_4$ as PV absorbers. Their structural, vibrational,
electronic, optical, excitonic, thermoelectric and photocatalytic properties are first
calculated with the GGA-PBEsol and HSE06 functionals. The calculated band gaps, effective
masses and optical constants are then used as inputs to finite-difference time-domain
(FDTD) optical simulations and finite-element drift--diffusion (CHARGE) electrical
simulations of FTO/IGZO/XGa$_2$Te$_4$/CuAlO$_2$/Pt cells, following the strategy of
linking first-principles and device-level modeling used for other emerging absorbers
\cite{bochaoui2025,sharma2025}. The device analysis addresses band alignment, optical
absorption and interference, the spatial distribution of photogeneration, and the
dependence of performance on absorber doping and thickness. Finally, we examine the
consistency between the first-principles inputs of the optical and electrical models,
because it determines how far the simulated device metrics can be interpreted
quantitatively.

\section{Materials and methodology}
\label{sec:methods}

\subsection{First-principles calculations of CdGa$_2$Te$_4$ and ZnGa$_2$Te$_4$}
\label{sec:dft-method}

First-principles calculations were performed with the plane-wave pseudopotential method
implemented in the Cambridge Serial Total Energy Package (CASTEP) \cite{clark2005}.
Ultrasoft pseudopotentials \cite{vanderbilt1990} were used for structural relaxation,
with the Cd $4d^{10}5s^{2}$, Zn $3d^{10}4s^{2}$, Ga $3d^{10}4s^{2}4p^{1}$ and
Te $5s^{2}5p^{4}$ states treated as valence electrons. Exchange and correlation were
treated within the generalized gradient approximation (GGA) \cite{perdew1996} using the
PBEsol functional \cite{perdew2008}. A plane-wave cut-off energy of 600~eV and an
$8\times8\times4$ Monkhorst--Pack $k$-point grid \cite{monkhorst1976} were used. Geometry
optimization was converged to $2\times10^{-6}$~eV~atom$^{-1}$ in total energy,
$5\times10^{-3}$~eV~\AA$^{-1}$ in residual force, 0.02~GPa in stress and
$5\times10^{-4}$~\AA{} in atomic displacement. Both compounds were optimized in the
tetragonal space group $I\bar{4}$ (No.~82); the resulting lattice parameters are listed in
Table~\ref{tab:lattice} and the unit cell is shown in Fig.~\ref{fig:structure}. Because
semilocal functionals underestimate band gaps, the band structures were also calculated
with the HSE06 screened hybrid functional \cite{heyd2003,krukau2006}, using a screening
parameter of $\omega=0.2$~\AA$^{-1}$, 25\% exact exchange and a $4\times4\times2$
$k$-point grid; as required by the CASTEP implementation of hybrid functionals,
norm-conserving pseudopotentials \cite{troullier1991} were used for these calculations.
The density of states, effective masses and frequency-dependent optical properties were
obtained at the PBEsol level. Phonon dispersion curves were calculated using
density-functional perturbation theory (DFPT) \cite{baroni2001} with norm-conserving
pseudopotentials on a $4\times4\times2$ $q$-point grid, followed by Fourier interpolation
along the high-symmetry path. Thermoelectric transport coefficients were obtained from the
PBEsol band structure by solving the semiclassical Boltzmann transport equation in the
constant relaxation time approximation (CRTA) and the rigid-band approximation, using the
BoltzTraP code \cite{madsen2006} with the $k$-mesh interpolated to five times the density
of the original grid.
\subsection{Device model and coupled FDTD--CHARGE workflow}
\label{sec:device-method}

\begin{figure*}[pos=t]
  \centering
  \includegraphics[width=0.85\textwidth]{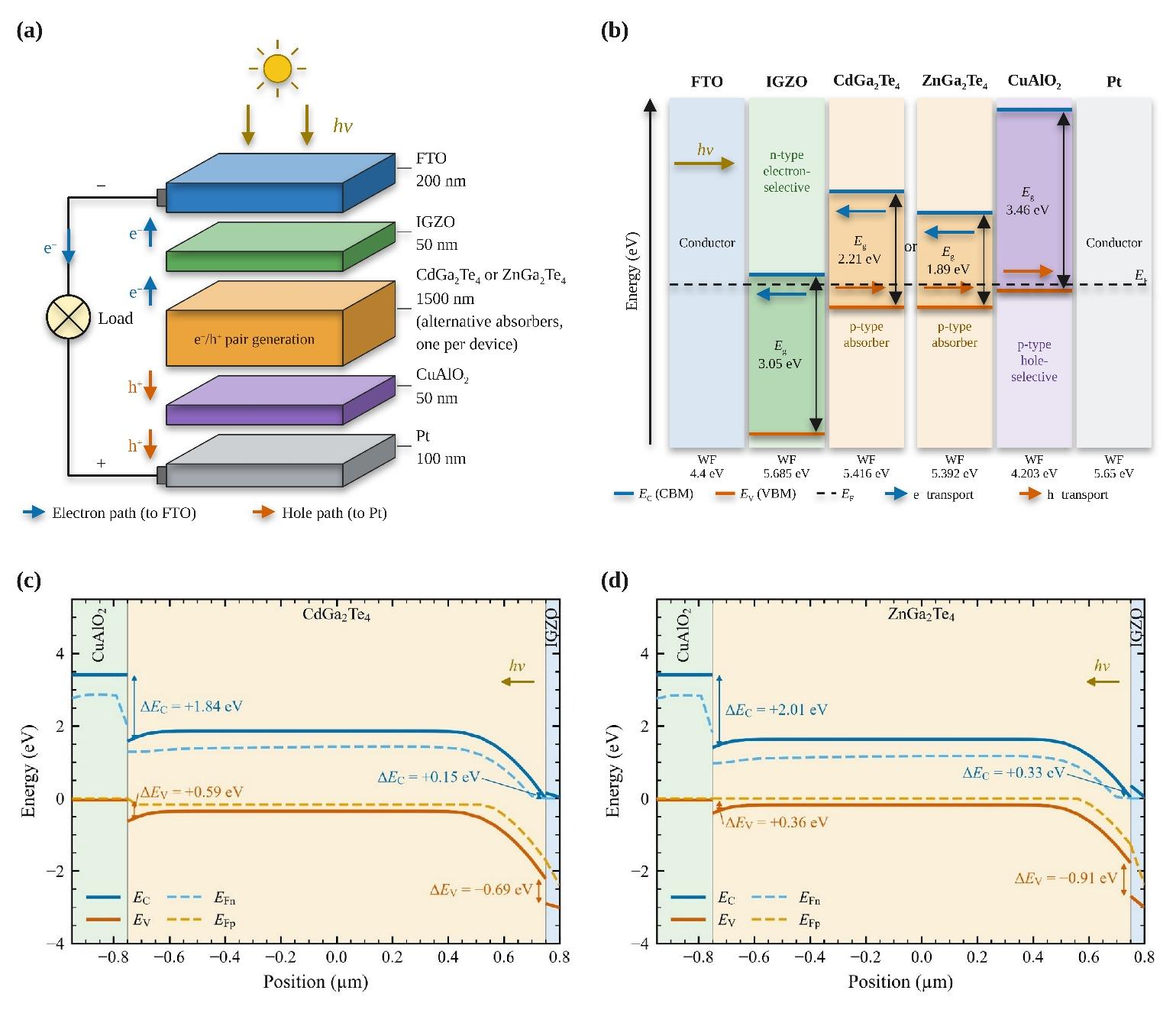}
  \caption{(a) Schematic of the FTO/IGZO/XGa$_2$Te$_4$/CuAlO$_2$/Pt (X = Cd, Zn) solar
           cell showing the layer thicknesses, the direction of illumination and the
           electron and hole extraction paths. (b) Energy-level diagram of the constituent
           layers showing the band gaps and the work functions (WF) used in the CHARGE
           model. (c, d) Simulated equilibrium energy-band profiles, with the electron and
           hole quasi-Fermi levels ($E_{\mathrm{Fn}}$, $E_{\mathrm{Fp}}$), across the
           (c) CdGa$_2$Te$_4$- and (d) ZnGa$_2$Te$_4$-based devices, including the band
           offsets at the CuAlO$_2$/absorber and absorber/IGZO interfaces.}
  \label{fig:device}
\end{figure*}

The photovoltaic potential of CdGa$_2$Te$_4$ and ZnGa$_2$Te$_4$ was assessed in a planar
FTO/IGZO/XGa$_2$Te$_4$/CuAlO$_2$/Pt (X = Cd, Zn) heterojunction, shown schematically in
Fig.~\ref{fig:device}(a). Light enters through a 200~nm FTO front electrode and a 50~nm
n-type IGZO ETL before it reaches the 1500~nm p-type XGa$_2$Te$_4$ absorber. A 50~nm
p-type CuAlO$_2$ HTL and a 100~nm Pt back electrode complete the stack. The two devices
differ only in the absorber, so any difference in their optical or electrical response
can be attributed to the absorber itself. The wide band gap of IGZO (3.05~eV) limits
parasitic absorption before light reaches the absorber, and the opposite doping of IGZO
($N_{\mathrm{D}} = 1\times10^{17}$~cm$^{-3}$) and CuAlO$_2$
($N_{\mathrm{A}} = 3\times10^{18}$~cm$^{-3}$) defines the electron- and hole-selective
contacts. The energy-level diagram in Fig.~\ref{fig:device}(b) and the simulated band
profiles in Fig.~\ref{fig:device}(c, d) identify the intended extraction pathways:
photogenerated electrons are transferred through IGZO to FTO, whereas holes are
transferred through CuAlO$_2$ to Pt.

Device modeling followed a sequential optical--electrical scheme in Ansys Lumerical
2024 R1, in which the FDTD solver and the finite-element
CHARGE solver were applied to the same two-dimensional (2D) device cross-section
\cite{firouzabadi2026,abdelhamied2026,wahba2023,roy2024}. The two solvers address
different physics and require different inputs. The FDTD solver is an electromagnetic
solver: it resolves the optical field in the full multilayer stack from the
wavelength-dependent complex refractive index of every layer (Table~\ref{tab:optical-inputs})
and returns the absorbed power and the spatial photogeneration profile under AM1.5G
illumination. It contains no carrier transport and therefore cannot by itself yield a
current density--voltage ($J$--$V$) characteristic. The CHARGE solver is an electrical
solver: it uses the electrical parameters in Table~\ref{tab:charge}, takes the FDTD
generation profile as the optical source term of the semiconductor equations, and solves
these self-consistently to obtain the band diagrams, carrier distributions and illuminated
$J$--$V$ characteristics. All photovoltaic figures of merit reported here therefore
originate from CHARGE. Coupling the solvers in this way captures thin-film interference,
the spatial non-uniformity of photogeneration and the transport and recombination losses
that a Beer--Lambert generation profile cannot resolve.

\subsubsection{Optical simulation with Lumerical FDTD}
\label{sec:fdtd-method}

The 2D FDTD solver discretizes Maxwell's curl equations on a Yee grid and propagates the
fields in the time domain; the frequency-domain fields are obtained by Fourier
transformation \cite{abdelhamied2026,wahba2023}. A broadband plane-wave source spanning
300--900~nm was launched at normal incidence onto the FTO surface, as indicated in
Fig.~\ref{fig:device}(a). Periodic boundary conditions were applied along the lateral
($x$) direction, consistent with a laterally invariant planar stack, and perfectly matched
layers (PMLs) were placed above the source and below the Pt electrode to absorb outgoing
radiation without spurious reflection \cite{abdelhamied2026,wahba2023}. The
auto-non-uniform mesh was used with a mesh accuracy of 3 and a minimum mesh
step of 1 nm, and to adequately resolve the electromagnetic field distribution across the different device layers. Frequency-domain power monitors above the source and below the Pt
electrode recorded the reflectance $R(\lambda)$ and transmittance $T(\lambda)$, from which
the total absorptance of the stack was obtained as
\begin{equation}
  A(\lambda) = 1 - R(\lambda) - T(\lambda).
  \label{eq:ART}
\end{equation}
Because the 100~nm Pt back electrode renders the stack effectively opaque, $T(\lambda)$ is
negligible and $A(\lambda)$ is set mainly by the balance between front-surface reflection
and absorption within the stack. $A(\lambda)$ also contains parasitic absorption in FTO,
IGZO, CuAlO$_2$ and Pt; the fraction converted into carriers inside the absorber is
resolved from the spatial generation analysis. A 2D frequency-domain field monitor covering the entire device recorded the electric field
$\mathbf{E}(\mathbf{r},\lambda)$, which is reported as the normalized amplitude
$|E|/|E_0|$, where $E_0$ is the amplitude of the incident field.

Solar illumination was represented by the AM1.5G reference spectrum of ASTM G173-03
\cite{astmg173}, which integrates to an irradiance of 100~mW~cm$^{-2}$ and is the standard
reference for rating terrestrial photovoltaic devices. The source amplitude was
normalized in the simulation, and the AM1.5G weighting was applied during post-processing,
so that every wavelength contributes to absorption and photogeneration in proportion to
its share of the solar spectrum.

The reliability of an FDTD calculation depends directly on the optical constants assigned
to each layer. Table~\ref{tab:optical-inputs} summarizes the function, thickness and source
of the $n(\lambda)$ and $k(\lambda)$ data for every layer. For CdGa$_2$Te$_4$ and
ZnGa$_2$Te$_4$, $n(\lambda)$ and $k(\lambda)$ were derived from the frequency-dependent
HSE06 dielectric function calculated in this work (Section~\ref{sec:dft-method}) through
Eqs.~(\ref{eq:n}) and (\ref{eq:k}). This follows the approach of transferring
first-principles optical constants into FDTD for materials that lack measured dispersion
data \cite{sharma2025}. Because the HSE06 functional accurately reproduces the
quasiparticle band gaps, the absorption edges of the imported $k(\lambda)$ spectra align
with the HSE06 gaps of 561~nm for CdGa$_2$Te$_4$ and 654~nm for ZnGa$_2$Te$_4$,
ensuring physically consistent onset wavelengths in the simulated photocurrent.
The optical constants of FTO, IGZO, CuAlO$_2$ and Pt were taken from
the sources listed in Table~\ref{tab:optical-inputs}. All tabulated data were imported into
the FDTD material database as sampled data and fitted with the multi-coefficient material
model to ensure stable broadband time-domain propagation.

\begin{table}[!htb]
  \centering
  \caption{Function, thickness and source of the optical constants $n(\lambda)$ and
           $k(\lambda)$ of each layer used in the FDTD simulations.}
  \label{tab:optical-inputs}
  \footnotesize
  \setlength{\tabcolsep}{3pt}
  \begin{tabular}{lp{2.5cm}cc}
    \toprule
    Layer & Function & $d$ (nm) & $n$, $k$ source \\
    \midrule
    FTO            & Transparent front electrode  & 200  & \cite{vonrottkay1996} \\
    IGZO           & Electron-transport layer     & 50   & \cite{talagrand2015} \\
    CdGa$_2$Te$_4$ & Absorber                     & 1500 & This work\\
    ZnGa$_2$Te$_4$ & Absorber                     & 1500 & This work\\
    CuAlO$_2$      & Hole-transport layer         & 50   & \cite{kang2026} \\
    Pt             & Back electrode and reflector & 100  & \cite{tselin2024} \\
    \bottomrule
  \end{tabular}
\end{table}

\subsubsection{Photogeneration rate and heat-source density}
\label{sec:gq-method}

The local absorbed power density per unit source intensity was obtained from the
divergence of the time-averaged Poynting vector, which for non-magnetic media reduces to
\cite{wahba2023,roy2024}
\begin{equation}
  P_{\mathrm{abs}}(\mathbf{r},\lambda) = \frac{1}{2}\,\omega\,\varepsilon_0\,
  \varepsilon''(\mathbf{r},\lambda)\,\big|\mathbf{E}(\mathbf{r},\lambda)\big|^{2},
  \label{eq:pabs}
\end{equation}
where $\omega$ is the angular frequency, $\varepsilon_0$ is the vacuum permittivity,
$\varepsilon'' = 2nk$ is the imaginary part of the relative permittivity and
$\mathbf{E}(\mathbf{r},\lambda)$ is the local electric field. Two distinct quantities were
derived from $P_{\mathrm{abs}}$. The first is the photogeneration rate $G(\mathbf{r})$, the
number of electron--hole pairs created per unit volume and time
(cm$^{-3}$~s$^{-1}$). Assuming that every absorbed photon generates one electron--hole
pair,
\begin{equation}
  G(\mathbf{r}) = \int_{\lambda_1}^{\lambda_2}
  \frac{P_{\mathrm{abs}}(\mathbf{r},\lambda)}{\hbar\omega}\,
  \frac{I_{\mathrm{AM1.5G}}(\lambda)}{I_0(\lambda)}\,\mathrm{d}\lambda ,
  \label{eq:gen}
\end{equation}
where $\hbar\omega$ is the photon energy, $I_{\mathrm{AM1.5G}}(\lambda)$ is the AM1.5G
spectral irradiance, $I_0(\lambda)$ is the source intensity used in the simulation, and
$\lambda_1 = 300$~nm and $\lambda_2 = 900$~nm bound the simulated spectral window.
Eq.~(\ref{eq:gen}) counts every photon absorbed in the absorber as one electron--hole pair,
whatever its energy relative to the band gap, and therefore assumes unit internal quantum
efficiency; transport and recombination losses are evaluated subsequently in CHARGE. The
resulting 2D generation map, $G(x,y)$, was exported in Lumerical's generation-rate format
for the electrical simulations.

The second quantity is the heat-source density $Q(\mathbf{r})$, an energy flow per unit
volume (W~cm$^{-3}$) that is not used by CHARGE but indicates where absorbed optical power
is deposited as heat. Two definitions are physically meaningful and must be distinguished.
If no carriers are extracted, all absorbed power is ultimately converted into heat, and
\begin{equation}
  Q_{\mathrm{abs}}(\mathbf{r}) = \int_{\lambda_1}^{\lambda_2}
  P_{\mathrm{abs}}(\mathbf{r},\lambda)\,
  \frac{I_{\mathrm{AM1.5G}}(\lambda)}{I_0(\lambda)}\,\mathrm{d}\lambda .
  \label{eq:qabs}
\end{equation}
If only the immediate thermalization of hot carriers to the band edges is counted, each
absorbed photon above the gap contributes the excess energy $\hbar\omega - E_g$, giving
\begin{equation}
  Q_{\mathrm{th}}(\mathbf{r}) = \int_{\hbar\omega > E_g}
  P_{\mathrm{abs}}(\mathbf{r},\lambda)
  \left(1-\frac{E_g}{\hbar\omega}\right)
  \frac{I_{\mathrm{AM1.5G}}(\lambda)}{I_0(\lambda)}\,\mathrm{d}\lambda .
  \label{eq:qth}
\end{equation}
Comparing Eqs.~(\ref{eq:gen}) and (\ref{eq:qabs}) shows that the ratio
$Q_{\mathrm{abs}}/(qG)$ equals the mean energy of the photons absorbed at $\mathbf{r}$ and
must therefore lie between 1.38 and 4.13~eV for the 300--900~nm window. This ratio is used
in Section~\ref{sec:genheat} to test which definition underlies the reported $Q$ maps,
which were obtained from the Lumerical heat-source analysis object using AM1.5G spectral weighting.

\subsubsection{Electrical simulation with Lumerical CHARGE}
\label{sec:charge-method}

Carrier transport was modeled with the CHARGE solver, which discretizes the semiconductor
equations on a finite-element mesh and solves them self-consistently under steady-state
conditions \cite{firouzabadi2026,abdelhamied2026,roy2024}. The 2D geometry and layer
thicknesses were identical to those of the FDTD model (Fig.~\ref{fig:device}(a)), and
$G(x,y)$ from Eq.~(\ref{eq:gen}) was imported as a spatially resolved source term. The
electrostatic potential $\varphi$ follows from Poisson's equation,
\begin{equation}
  \nabla\cdot\left(\varepsilon_0\varepsilon_r\nabla\varphi\right)
  = -q\left(p - n + N_{\mathrm{D}}^{+} - N_{\mathrm{A}}^{-}\right),
  \label{eq:poisson}
\end{equation}
where $\varepsilon_r$ is the relative permittivity, $q$ is the elementary charge, $n$ and
$p$ are the electron and hole densities, and $N_{\mathrm{D}}^{+}$ and $N_{\mathrm{A}}^{-}$
are the ionized donor and acceptor densities. The electron and hole current densities are
described by the drift--diffusion model \cite{firouzabadi2026},
\begin{align}
  \mathbf{J}_n &= q\mu_n n\mathbf{E} + qD_n\nabla n ,
  \label{eq:jn} \\
  \mathbf{J}_p &= q\mu_p p\mathbf{E} - qD_p\nabla p ,
  \label{eq:jp}
\end{align}
where $\mathbf{E} = -\nabla\varphi$ is the electrostatic field, $\mu_n$ and $\mu_p$ are the
carrier mobilities, and the diffusion coefficients follow the Einstein relation,
$D_{n,p} = \mu_{n,p}k_{\mathrm{B}}T/q$. Charge conservation is enforced by the continuity
equations
\begin{align}
  \frac{\partial n}{\partial t} &= \frac{1}{q}\nabla\cdot\mathbf{J}_n + G - R ,
  \label{eq:cont-n} \\
  \frac{\partial p}{\partial t} &= -\frac{1}{q}\nabla\cdot\mathbf{J}_p + G - R ,
  \label{eq:cont-p}
\end{align}
whose left-hand sides vanish at steady state; $G$ is the imported optical generation rate
and $R$ the net recombination rate. Non-radiative recombination through bulk defect states
was described by the Shockley--Read--Hall (SRH) model \cite{roy2024},
\begin{equation}
  R_{\mathrm{SRH}} = \frac{np - n_i^{2}}
  {\tau_p\left(n + n_1\right) + \tau_n\left(p + p_1\right)} ,
  \label{eq:srh}
\end{equation}
where $n_i$ is the intrinsic carrier density, $n_1$ and $p_1$ are the electron and hole
densities when the Fermi level coincides with the trap level, and $\tau_n$ and $\tau_p$
are the electron and hole lifetimes. SRH recombination was the only bulk recombination
mechanism enabled; radiative and Auger recombination were not included. Recombination at
the IGZO/absorber and absorber/CuAlO$_2$ interfaces was described by surface recombination
velocities of 2612~cm~s$^{-1}$ for electrons and 1238~cm~s$^{-1}$ for holes.

The layer parameters used in CHARGE are listed in Table~\ref{tab:charge}. The band gaps and
carrier effective masses of the absorbers were taken from the first-principles
calculations of this work, which links the device model to the calculated electronic
structure; the remaining parameters were taken from the sources listed in
Table~\ref{tab:charge}. The work function assigned to each semiconductor in CHARGE refers
to the undoped material and fixes the band alignment; the band offsets reported in
Section~\ref{sec:alignment} follow from these work functions and band gaps. Both absorbers
were assigned an acceptor density of $2\times10^{16}$~cm$^{-3}$ and SRH electron and hole
lifetimes of 1~ns. The short minority-carrier (electron) lifetime is a conservative choice for
emerging absorbers whose defect chemistry has not yet been established experimentally.
FTO and Pt were treated as conductors serving as the cathode and anode, respectively, with
contact behavior set by their work functions of 4.4 and 5.65~eV. All simulations were
performed at 300~K.

\begin{table*}[!t]
  \centering
  \caption{Electrical parameters of each layer used in the Lumerical CHARGE simulations.}
  \label{tab:charge}
  \footnotesize
  \setlength{\tabcolsep}{4pt}
  \begin{tabular}{lcccccc}
    \toprule
    Parameter & FTO & IGZO & CdGa$_2$Te$_4$ & ZnGa$_2$Te$_4$ & CuAlO$_2$ & Pt \\
    \midrule
    Layer thickness (nm)                          & 200       & 50            & 1500          & 1500          & 50            & 100 \\
    Material type                                 & Conductor & Semiconductor & Semiconductor & Semiconductor & Semiconductor & Conductor \\
    Band gap, $E_g$ (eV)                          & ---       & 3.05          & 2.21$^{a}$          & 1.895$^{a}$    & 3.46          & --- \\
    Work function, WF (eV)                        & 4.4       & 5.685         & 5.416         & 5.392         & 4.203         & 5.65 \\
    Electron mobility, $\mu_n$ (cm$^2$~V$^{-1}$~s$^{-1}$) & --- & 15        & 15.178        & 17.571        & 2             & --- \\
    Hole mobility, $\mu_p$ (cm$^2$~V$^{-1}$~s$^{-1}$)     & --- & 0.1       & 11.249        & 17.648        & 8.6           & --- \\
    Donor density, $N_{\mathrm{D}}$ (cm$^{-3}$)    & ---       & $1\times10^{17}$ & ---        & ---           & ---           & --- \\
    Acceptor density, $N_{\mathrm{A}}$ (cm$^{-3}$) & ---       & ---           & $2\times10^{16}$ & $2\times10^{16}$ & $3\times10^{18}$ & --- \\
    Electron effective mass, $m_n^*/m_0$          & ---       & 0.341         & 0.1997        & 0.4001        & 0.197         & --- \\
    Hole effective mass, $m_p^*/m_0$              & ---       & 0.341         & 0.8993        & 0.8013        & 0.801         & --- \\
    Relative permittivity, $\varepsilon_r$        & ---       & 10            & 10.8          & 10.9          & 60            & --- \\
    SRH electron lifetime, $\tau_n$ (s)           & ---       & $1\times10^{-7}$ & $1\times10^{-9}$ & $1\times10^{-9}$ & $1\times10^{-7}$ & --- \\
    SRH hole lifetime, $\tau_p$ (s)               & ---       & $1\times10^{-7}$ & $1\times10^{-9}$ & $1\times10^{-9}$ & $1\times10^{-7}$ & --- \\
    Reference                                     & \cite{hossain2024sr} & \cite{hossain2023sr} & This work & This work & \cite{hossain2024sr} & \cite{derry1989work} \\
    \bottomrule
  \end{tabular}

  \vspace{2pt}
  \begin{minipage}{0.95\textwidth}
  \footnotesize $^{a}$~The HSE06 gap of ZnGa$_2$Te$_4$ is 2.21~eV and ZnGa$_2$Te$_4$ is 1.895~eV
  (Table~\ref{tab:gaps}).
  \end{minipage}
\end{table*}

\subsubsection{Extraction of photovoltaic parameters and parametric studies}
\label{sec:pv-method}

Illuminated $J$--$V$ characteristics were obtained in CHARGE by sweeping the anode bias
from 0 to 2~V in steps of 0.02~V, beyond the open-circuit point of
each device. The short-circuit current density ($J_{\mathrm{SC}}$) and $V_{\mathrm{OC}}$
were extracted from the intercepts of the $J$--$V$ curve, and the maximum power point
($J_{\mathrm{mpp}}$, $V_{\mathrm{mpp}}$) was located from the power density--voltage
($P$--$V$) curve. The fill factor (FF) and PCE were then calculated as
\cite{abdelhamied2026,karna2023}
\begin{align}
  \mathrm{FF} &= \frac{P_{\max}}{J_{\mathrm{SC}}V_{\mathrm{OC}}}
               = \frac{J_{\mathrm{mpp}}V_{\mathrm{mpp}}}{J_{\mathrm{SC}}V_{\mathrm{OC}}} ,
  \label{eq:ff} \\
  \mathrm{PCE} &= \frac{J_{\mathrm{SC}}V_{\mathrm{OC}}\,\mathrm{FF}}{P_{\mathrm{in}}}
                  \times 100\% ,
  \label{eq:pce}
\end{align}
where $P_{\max}$ is the maximum output power density and
$P_{\mathrm{in}} = 100$~mW~cm$^{-2}$ is the incident AM1.5G power density. To establish
the design window of both architectures, the absorber acceptor density was varied from
$1\times10^{14}$ to $1\times10^{18}$~cm$^{-3}$, and the absorber thickness from 0.4 to
2.0~$\mu$m in 0.2~$\mu$m steps, while all other parameters were held at the values in
Tables~\ref{tab:optical-inputs} and \ref{tab:charge}. For each absorber thickness, the FDTD
simulation was repeated and the corresponding generation profile was re-imported into
CHARGE, so that thickness-dependent interference and absorption were captured
consistently.

\section{Results and discussion}
\label{sec:results}

\subsection{Structural properties and dynamic stability}
\label{sec:structure}

\begin{figure*}[pos=t]
  \centering
  \includegraphics[width=0.85\textwidth]{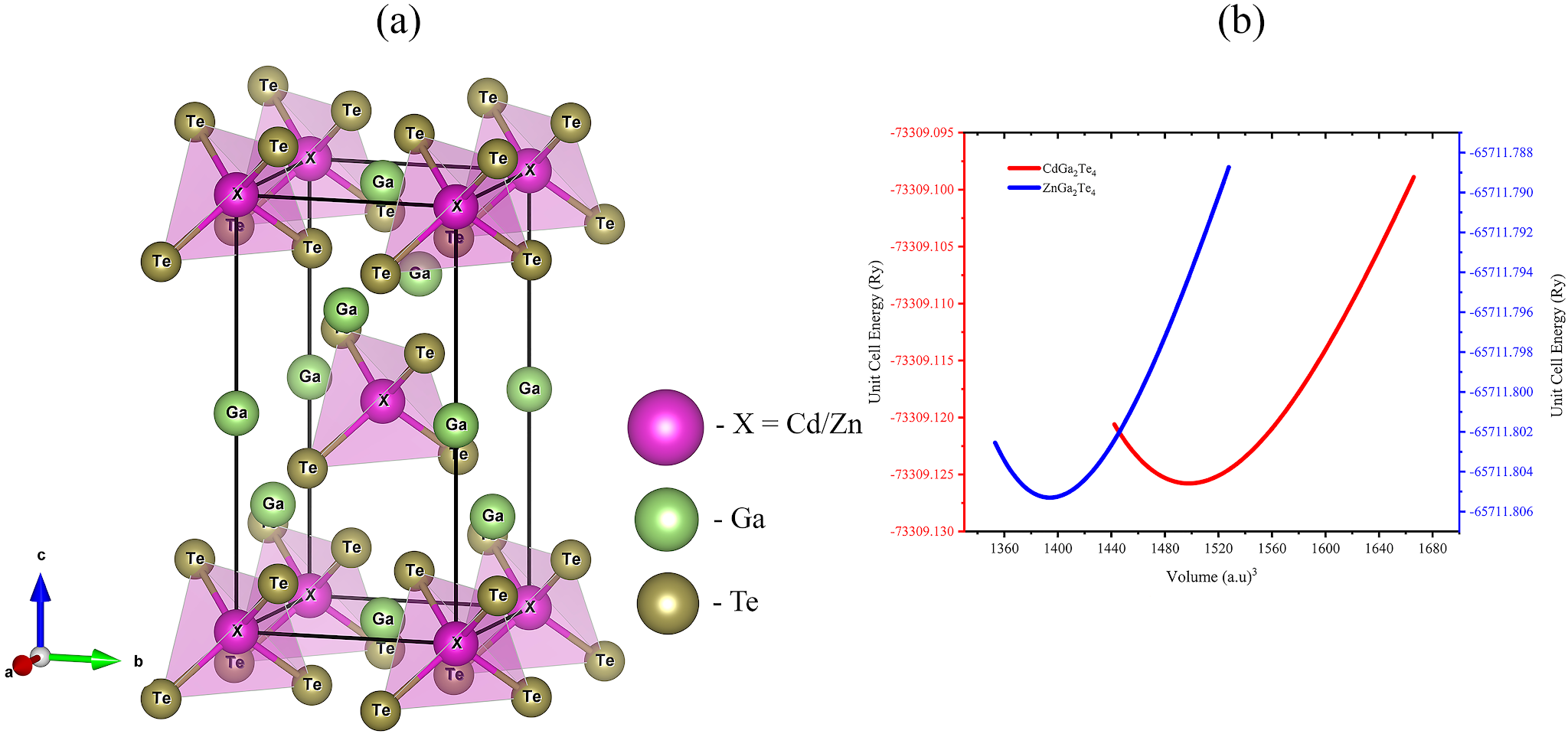}
  \caption{(a) Optimized crystal structure and (b) total energy as a function of
           unit-cell volume ($E$--$V$) of XGa$_2$Te$_4$ (X = Cd, Zn).}
  \label{fig:structure}
\end{figure*}

CdGa$_2$Te$_4$ and ZnGa$_2$Te$_4$ crystallize in the defect-chalcopyrite structure with the
tetragonal space group $I\bar{4}$ (No.~82). Fig.~\ref{fig:structure}(a) shows the
optimized unit cell, visualized with VESTA \cite{momma2008}. The X = Cd or Zn atoms occupy
the 2a Wyckoff site (0, 0, 0), Ga atoms occupy the 2b (1/2, 1/2, 0) and 2d (1/2, 0, 1/4)
sites, and Te atoms occupy the 8g site (Table~\ref{tab:wyckoff}). The ground-state
structures were obtained by minimizing the total energy at zero temperature and pressure;
the lattice parameters and the formation and cohesive energies are listed in
Table~\ref{tab:lattice}. For CdGa$_2$Te$_4$, the calculated lattice constants
$a = 6.22$~\AA{} and $c = 11.92$~\AA{} deviate from the experimental values of 6.04 and
12.135~\AA{} \cite{jahangirli2020} by $+3.0\%$ and $-1.8\%$, respectively. The equilibrium
volume $V_0$ was obtained by fitting the total energy as a function of volume
(Fig.~\ref{fig:structure}(b)) to the Murnaghan equation of state \cite{tyuterev2006},
\begin{equation}
  \begin{split}
  E(V) = E_0 &+ \frac{B_0 V}{B_0'\left(B_0'-1\right)} \\
             &\times\left[B_0'\left(1-\frac{V_0}{V}\right)
              + \left(\frac{V_0}{V}\right)^{B_0'} - 1\right],
  \end{split}
  \label{eq:bm}
\end{equation}
where $E_0$ is the minimum energy, $B_0$ is the bulk modulus and $B_0'$ is its pressure
derivative. The fit gives $B_0 = 38.2$~GPa and $B_0' = 4.8$ for
CdGa$_2$Te$_4$, and $B_0 = 44.1$~GPa and $B_0' = 4.6$ for
ZnGa$_2$Te$_4$.

\begin{table}[!htb]
  \centering
  \caption{Wyckoff positions of the different atomic species in the XGa$_2$Te$_4$
           (X = Cd, Zn) crystal structures.}
  \label{tab:wyckoff}
  \small
  \setlength{\tabcolsep}{3pt}   
  \begin{tabular}{llllll}
    \toprule
    Compound & Wyckoff & Atom & $x$ & $y$ & $z$ \\
    \midrule
    \multirow{4}{*}{CdGa$_2$Te$_4$}
      & 2a & Cd & 0        & 0        & 0        \\
      & 2b & Ga & 1/2      & 1/2      & 0        \\
      & 2d & Ga & 1/2      & 0        & 1/4      \\
      & 8g & Te & 0.226747 & 0.755192 & 0.362783 \\
    \midrule
    \multirow{4}{*}{ZnGa$_2$Te$_4$}
      & 2a & Zn & 0        & 0        & 0        \\
      & 2b & Ga & 1/2      & 1/2      & 0        \\
      & 2d & Ga & 1/2      & 0        & 1/4      \\
      & 8g & Te & 0.761913 & 0.265063 & 0.132453 \\
    \bottomrule
  \end{tabular}
\end{table}

\begin{table*}[!t]
  \centering
  \caption{Lattice parameters, total energy, formation energy ($\Delta E_f$) and cohesive
           energy ($\Delta E_c$) of the XGa$_2$Te$_4$ (X = Cd, Zn) structures.}
  \label{tab:lattice}
  \small
  \setlength{\tabcolsep}{4pt}   
  \begin{tabular}{lccccccl}
    \toprule
    Compound & Total energy (eV) & $a$ (\AA) & $c$ (\AA) & $c/a$
             & $\Delta E_f$ (eV/atom) & $\Delta E_c$ (eV/atom) & Ref. \\
    \midrule
    CdGa$_2$Te$_4$ & $-16075.038$ & 6.22  & 11.92  & 1.92 & $-1.44$ & $-3.634$ & This work \\
    ZnGa$_2$Te$_4$ & $-17210.346$ & 6.014 & 11.92  & 1.98 & $-1.39$ & $-3.756$ & This work \\
    CdGa$_2$Te$_4$ & ---          & 6.04  & 12.135 & 2.01 & ---     & ---      & Exp.\ \cite{jahangirli2020} \\
    \bottomrule
  \end{tabular}
\end{table*}

The energetic stability of the compounds was assessed through the formation energy
$\Delta E_f$ and the cohesive energy $\Delta E_c$, calculated from
Eqs.~(\ref{eq:formation}) and (\ref{eq:cohesive}):
\begin{align}
  \Delta E_f(\text{XGa}_2\text{Te}_4) &= \frac{1}{14}\Big[E_{\mathrm{total}}
      - \big(2E_{\mathrm{gr}}(\text{X}) \nonumber \\
  &\qquad + 4E_{\mathrm{gr}}(\text{Ga}) + 8E_{\mathrm{gr}}(\text{Te})\big)\Big] ,
  \label{eq:formation} \\[4pt]
  \Delta E_c(\text{XGa}_2\text{Te}_4) &= \frac{1}{14}\Big[E_{\mathrm{total}}
      - \big(2E_{\mathrm{iso}}(\text{X}) \nonumber \\
  &\qquad + 4E_{\mathrm{iso}}(\text{Ga}) + 8E_{\mathrm{iso}}(\text{Te})\big)\Big] ,
  \label{eq:cohesive}
\end{align}
where $E_{\mathrm{total}}$ is the total energy of the conventional cell containing two X,
four Ga and eight Te atoms, $E_{\mathrm{gr}}$ is the energy per atom of each element in its
most stable bulk phase, and $E_{\mathrm{iso}}$ is the energy of the isolated atom in
vacuum. The negative formation energies in Table~\ref{tab:lattice} indicate that both
compounds are stable with respect to decomposition into their elements, and the negative
cohesive energies confirm that both crystals are strongly bound. Stability against the
elements is a necessary but not sufficient condition for synthesizability, which also
requires stability against competing phases such as CdTe, ZnTe and Ga$_2$Te$_3$; a
convex-hull analysis would settle this point.

\begin{figure*}[pos=t]
  \centering
  \includegraphics[width=0.85\textwidth]{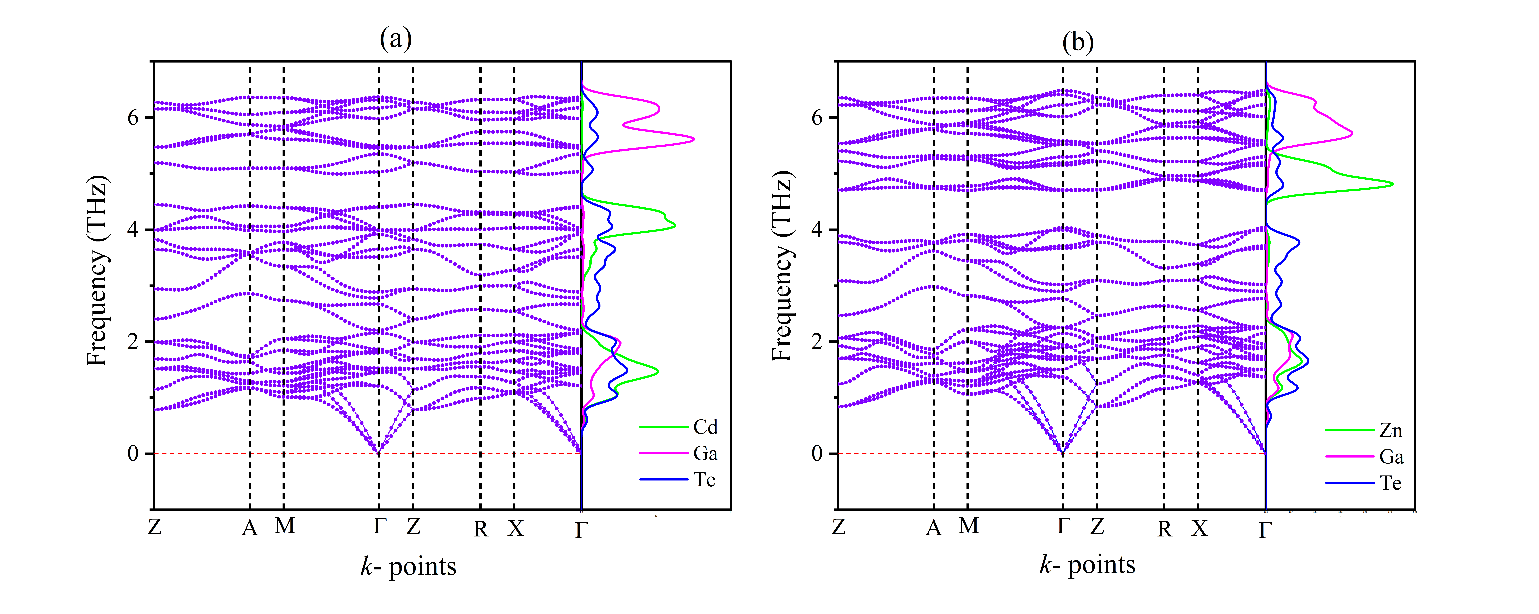}
  \caption{Calculated phonon dispersion (left) and phonon partial density of states
           (right) of (a) CdGa$_2$Te$_4$ and (b) ZnGa$_2$Te$_4$ along the
           Z--A--M--$\Gamma$--Z--R--X--$\Gamma$ path.}
  \label{fig:phonon}
\end{figure*}

Phonons, the quanta of lattice vibrations, govern the dynamic stability, thermal transport
and electron--phonon interactions of a crystal. Their spectrum is set by the atomic
masses, the interatomic force constants and the crystal symmetry
\cite{bouhenna2025,rahman2024,zaman2021}, so phonon dispersion curves reveal dynamic
stability, possible phase transitions and the likely magnitude of the lattice thermal
conductivity. Fig.~\ref{fig:phonon} shows the phonon dispersion along the
Z--A--M--$\Gamma$--Z--R--X--$\Gamma$ path and the corresponding phonon partial density of
states (PDOS). No imaginary frequencies occur anywhere in the Brillouin zone, which
confirms that both structures are dynamically stable. The three acoustic branches vanish
linearly at $\Gamma$ and extend only to $\approx$1.5--2~THz, and the whole spectrum lies
below $\approx$6.5~THz. The PDOS shows that the heavy Te atoms dominate the acoustic and
low-lying optical branches. Cd contributes mainly near $\approx$1.5 and $\approx$4.1~THz,
whereas the lighter Zn contributes near $\approx$4.8~THz, and Ga dominates the highest
optical branches between $\approx$5 and 6.5~THz. A small gap separates the upper optical
branches at $\approx$4.5--5.0~THz in CdGa$_2$Te$_4$ and $\approx$4.0--4.6~THz in
ZnGa$_2$Te$_4$. Low acoustic cut-off frequencies and a compact phonon spectrum generally
imply low phonon group velocities and hence a low lattice thermal conductivity; this point
is revisited qualitatively in Section~\ref{sec:thermo}, because the lattice thermal
conductivity was not calculated here.

\subsection{Electronic properties}
\label{sec:electronic}

\subsubsection{Band structures and density of states}
\label{sec:bands}

\begin{figure*}[pos=t]
  \centering
  \includegraphics[width=0.8\textwidth]{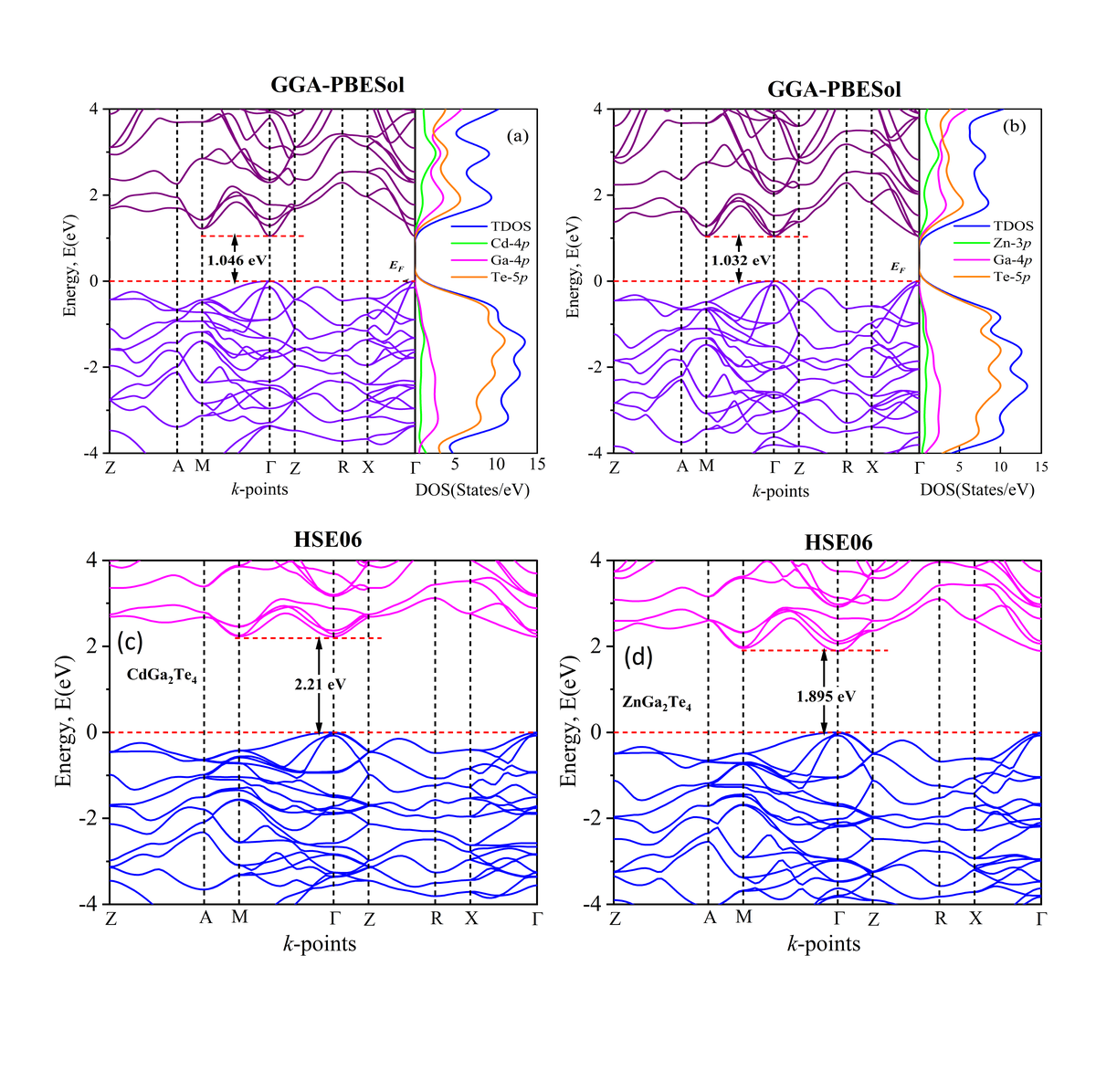}
  \caption{Electronic band structures and the corresponding total and partial density of
           states of CdGa$_2$Te$_4$ (a, c) and ZnGa$_2$Te$_4$ (b, d)
           calculated with the GGA-PBEsol (a, b) and HSE06 (c, d) functionals.}
  \label{fig:bands-hse}
\end{figure*}

The electronic band structure determines the optical response, the carrier effective
masses and the transport properties of a semiconductor, and it underpins materials design
for applications ranging from catalysis to superconductivity
\cite{norskov2009,hautier2010,calandra2007}. Figs.~\ref{fig:bands-hse}(a) and
\ref{fig:bands-hse}(b) show the band structures of CdGa$_2$Te$_4$ and ZnGa$_2$Te$_4$ calculated
with GGA-PBEsol along the high-symmetry directions of the first Brillouin zone, between
$-4$ and $+4$~eV. In both compounds, the valence-band maximum (VBM) and conduction-band
minimum (CBM) are located at $\Gamma$, giving direct band gaps,
\begin{equation}
  E_g = E_{\mathrm{CBM}} - E_{\mathrm{VBM}} ,
  \label{eq:bandgap}
\end{equation}
of 1.046~eV (CdGa$_2$Te$_4$) and 1.032~eV (ZnGa$_2$Te$_4$). Because semilocal functionals
systematically underestimate band gaps, the band structures were recalculated with HSE06
(Fig.~\ref{fig:bands-hse}), which gives larger direct gaps of 2.21 and 1.895~eV,
respectively, while preserving the band topology near $\Gamma$. The HSE06 values are
therefore used wherever the absolute gap matters, namely for the band-edge positions in
Section~\ref{sec:photocatalytic} and the device simulations in Section~\ref{sec:pv}.

The carrier effective masses, which control transport and the excitonic properties, were
obtained from the curvature of the band dispersion near the CBM and VBM according to
\begin{equation}
  \frac{1}{m^*} = \frac{1}{\hbar^{2}}\,\frac{\partial^{2}E(k)}{\partial k^{2}} ,
  \label{eq:mass}
\end{equation}
where $E(k)$ is the band energy as a function of the wave vector $k$ and $\hbar$ is the
reduced Planck constant. The effective densities of states of the conduction and valence
bands follow from the corresponding masses as
\begin{equation}
  N_{c,v} = 2\left(\frac{2\pi m_{e,h}^{*}k_{\mathrm{B}}T}{h^{2}}\right)^{3/2} ,
  \label{eq:Nc}
\end{equation}
evaluated at $T = 300$~K.

\begin{table*}[!t]
  \centering
  \caption{Electron ($m_e^*$) and hole ($m_h^*$) effective masses, effective densities of
           states of the conduction ($N_c$) and valence ($N_v$) bands, exciton binding
           energy ($E_b$), exciton Bohr radius ($R_{ex}$) and exciton dissociation
           temperature ($T_{ex}$) of CdGa$_2$Te$_4$ and ZnGa$_2$Te$_4$.}
  \label{tab:mass}
  \small
  \begin{tabular}{lccccccc}
    \toprule
    Compound & $m_e^*$ & $m_h^*$ & $N_c$ ($10^{18}$~cm$^{-3}$) & $N_v$ ($10^{19}$~cm$^{-3}$)
             & $E_b$ (meV) & $R_{ex}$ (\AA) & $T_{ex}$ (K) \\
    \midrule
    CdGa$_2$Te$_4$ & $0.214\,m_0$ & $-0.89\,m_0$ & 2.24 & 2.14 & 18.85 & 34.3 & 218.7 \\
    ZnGa$_2$Te$_4$ & $0.39\,m_0$  & $-0.85\,m_0$ & 6.35 & 1.80 & 26.81 & 23.0 & 311.1 \\
    \bottomrule
  \end{tabular}

  \vspace{2pt}
  \begin{minipage}{0.95\textwidth}
  \footnotesize The negative sign of $m_h^*$ reflects the downward curvature of the valence
  band; magnitudes are used in Eqs.~(\ref{eq:Nc}) and (\ref{eq:Eb}). $N_c$ and $N_v$ were
  evaluated at 300~K with Eq.~(\ref{eq:Nc}).
  \end{minipage}
\end{table*}

The effective masses in Table~\ref{tab:mass} were obtained by parabolic fitting of the
PBEsol bands near the CBM and VBM \cite{islam2020} and are given in units of the
free-electron mass $m_0$. CdGa$_2$Te$_4$ has a lighter electron mass ($0.214\,m_0$) than
ZnGa$_2$Te$_4$ ($0.39\,m_0$), whereas the hole masses are similar (0.89 and $0.85\,m_0$).
For a given scattering time, a lighter mass implies a higher mobility, so CdGa$_2$Te$_4$ is
expected to show the more favorable electron transport \cite{sze2007}.
Because electrons are the minority carriers in the p-type absorbers considered in
Section~\ref{sec:pv}, this difference is directly relevant to carrier collection. The hole
masses are about four (CdGa$_2$Te$_4$) and two (ZnGa$_2$Te$_4$) times larger than the
electron masses, indicating flatter valence bands, which is also reflected in the
effective valence-band density of states $N_v$ being roughly an order of magnitude larger
than $N_c$.

The density of states (DOS) gives the number of electronic states per unit energy and
thereby controls the carrier distribution and the strength of optical transitions. The
total (TDOS) and partial (PDOS) densities of states calculated with GGA-PBEsol are shown in
Fig.~\ref{fig:bands-hse}, with the Fermi level at zero energy. The TDOS vanishes at the Fermi
level, consistent with the semiconducting character found in the band structures, and the
gaps obtained from the TDOS agree with those from the band dispersions. The PDOS shows that
Te-5p states dominate both the upper valence band and the lower conduction band, with
smaller contributions from Ga-4p and Cd- or Zn-derived states. Replacing Cd with Zn alters the TDOS and PDOS only
slightly, because the Zn contribution near the band edges is small, which is consistent
with the small difference between the two band gaps. The strong hybridization among Te, Ga and Cd
(or Zn) states indicates a substantial covalent contribution to bonding.

Table~\ref{tab:gaps} compares the present band gaps with earlier DFT results. Our PBEsol
gap of ZnGa$_2$Te$_4$ agrees with the PBE value of Govindaraj \emph{et al.}
\cite{govindaraj2022}, and both the present HSE06 gaps and the mBJ gaps of Kumar
\emph{et al.} \cite{kumar2017} show the same trend of a larger gap for CdGa$_2$Te$_4$
than for ZnGa$_2$Te$_4$.

\begin{table*}[!t]
  \centering
  \caption{Band gaps of CdGa$_2$Te$_4$ and ZnGa$_2$Te$_4$ from the present and previous DFT
           studies, illustrating the consistent Cd~$>$~Zn gap trend.}
  \label{tab:gaps}
  \small
  \begin{tabular}{lllll}
    \toprule
    DFT study (year) & Material & Functional & Band gap (eV) & Type \\
    \midrule
    \multirow{4}{*}{Present work}
        & ZnGa$_2$Te$_4$ & GGA-PBEsol & 1.032 & Direct \\
        & CdGa$_2$Te$_4$ & GGA-PBEsol & 1.046 & Direct \\
        & ZnGa$_2$Te$_4$ & HSE06      & 1.895 & Direct \\
        & CdGa$_2$Te$_4$ & HSE06      & 2.21  & Direct \\
    \midrule
    \multirow{2}{*}{Kumar \emph{et al.} (2017) \cite{kumar2017}}
        & ZnGa$_2$Te$_4$ & mBJ & 1.61 & Direct \\
        & CdGa$_2$Te$_4$ & mBJ & 1.78 & Direct \\
    \midrule
    Govindaraj \emph{et al.} (2022) \cite{govindaraj2022}
        & ZnGa$_2$Te$_4$ & GGA-PBE & 1.01 & Direct \\
    \bottomrule
  \end{tabular}
\end{table*}

\subsection{Optical characteristics}
\label{sec:optical}

\begin{figure*}[pos=t]
  \centering
  \includegraphics[width=0.75\textwidth]{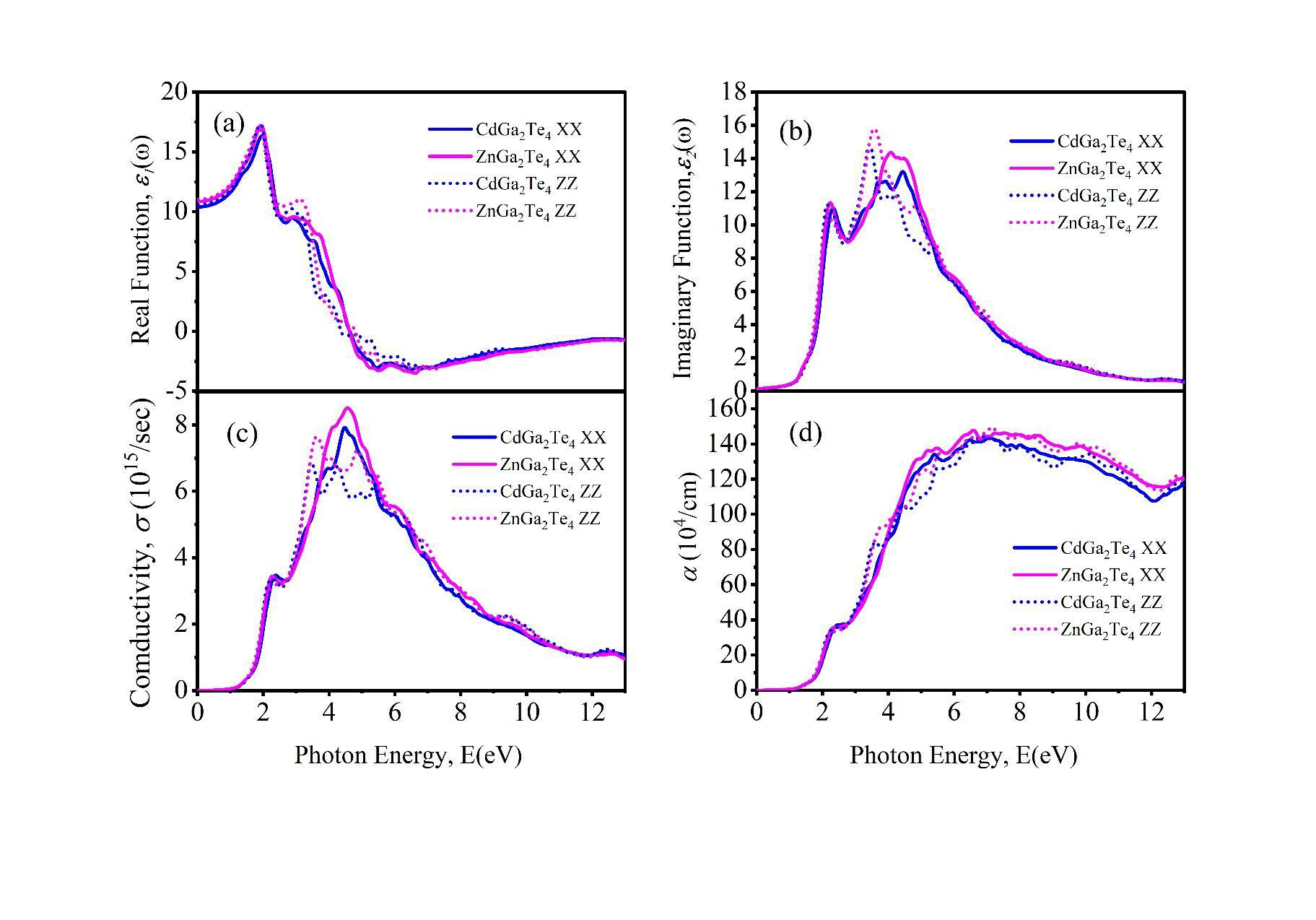}
  \caption{Polarization-dependent (a) real part of the dielectric function, (b) imaginary
           part of the dielectric function, (c) optical conductivity and (d) absorption
           coefficient of CdGa$_2$Te$_4$ and ZnGa$_2$Te$_4$ for light polarized along
           [100] and [001], calculated with GGA-PBEsol.}
  \label{fig:optical1}
\end{figure*}

\begin{figure*}[pos=t]
  \centering
  \includegraphics[width=0.75\textwidth]{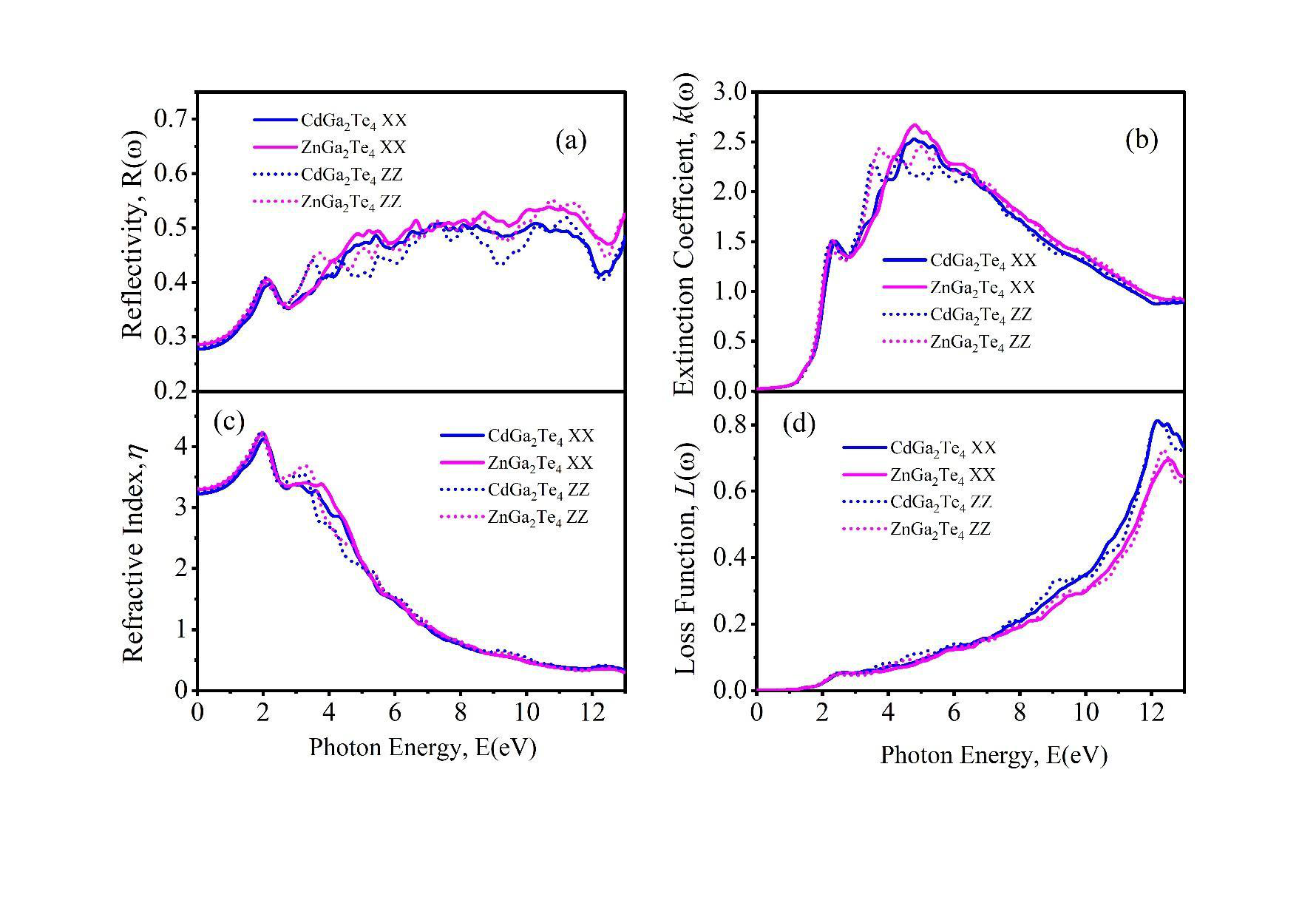}
  \caption{Polarization-dependent (a) reflectivity, (b) extinction coefficient,
           (c) refractive index and (d) energy-loss function of CdGa$_2$Te$_4$ and
           ZnGa$_2$Te$_4$, calculated with GGA-PBEsol.}
  \label{fig:optical2}
\end{figure*}

The optical response of a semiconductor is governed by interband transitions between
occupied and empty states and is described by the complex dielectric function
\begin{equation}
  \varepsilon(\omega) = \varepsilon_1(\omega) + i\,\varepsilon_2(\omega) ,
  \label{eq:dielectric}
\end{equation}
where $\varepsilon_1(\omega)$ and $\varepsilon_2(\omega)$ are the real and imaginary parts
and $\omega$ is the angular frequency of the incident light. $\varepsilon_2(\omega)$ was
calculated from the momentum matrix elements between occupied and unoccupied PBEsol
states, and $\varepsilon_1(\omega)$ was obtained from $\varepsilon_2(\omega)$ through the
Kramers--Kronig relation. The refractive index $n(\omega)$, extinction coefficient
$k(\omega)$, reflectivity $R(\omega)$, absorption coefficient $\alpha(\omega)$, energy-loss
function $L(\omega)$ and optical conductivity $\sigma(\omega)$ follow from
$\varepsilon(\omega)$ as \cite{rifat2026,liu2010}
\begin{align}
  n(\omega) &= \sqrt{\frac{|\varepsilon(\omega)| + \varepsilon_1(\omega)}{2}} ,
  \label{eq:n} \\
  k(\omega) &= \sqrt{\frac{|\varepsilon(\omega)| - \varepsilon_1(\omega)}{2}} ,
  \label{eq:k} \\
  R(\omega) &= \frac{(n-1)^2 + k^2}{(n+1)^2 + k^2} ,
  \label{eq:R} \\
  \alpha(\omega) &= \frac{2k\omega}{c} = \frac{4\pi k}{\lambda} ,
  \label{eq:alpha} \\
  L(\omega) &= \operatorname{Im}\!\left(\frac{-1}{\varepsilon(\omega)}\right)
             = \frac{\varepsilon_2(\omega)}{\varepsilon_1^2(\omega) + \varepsilon_2^2(\omega)} ,
  \label{eq:loss} \\
  \sigma(\omega) &= \sigma_1(\omega) + i\,\sigma_2(\omega),
  \quad \sigma_1(\omega) = \frac{\omega\,\varepsilon_2(\omega)}{4\pi} ,
  \label{eq:sigma}
\end{align}
where $c$ is the speed of light and $\sigma_1$ is written in Gaussian units (s$^{-1}$),
equivalent to $\varepsilon_0\omega\varepsilon_2$ in SI units. Figs.~\ref{fig:optical1} and
\ref{fig:optical2} show the optical spectra of both compounds for light polarized along
the [100] and [001] directions. All spectra were calculated with PBEsol, so their onsets
lie at the PBEsol gaps of $\approx$1.05~eV; with the HSE06 gaps, the onsets would shift to
$\approx$2.2~eV (CdGa$_2$Te$_4$) and $\approx$1.9~eV (ZnGa$_2$Te$_4$).

Figs.~\ref{fig:optical1}(a) and \ref{fig:optical1}(b) show $\varepsilon_1(\omega)$ and
$\varepsilon_2(\omega)$. The real part describes the polarization response and dispersion,
whereas the imaginary part describes absorption. The two polarization directions give
nearly identical spectra, so the optical response is close to isotropic. The static
dielectric constant $\varepsilon_1(0)$ is $\approx$10.5--11 for both compounds.
$\varepsilon_1(\omega)$ rises to a maximum of $\approx$17 near 2~eV, then decreases and
becomes negative above $\approx$4.6--5~eV, where the reflectivity is high. The absorptive
part $\varepsilon_2(\omega)$ has its onset near 1~eV, a first peak of $\approx$11 at
$\approx$2.3~eV and a broad maximum of $\approx$13--16 between 3.6 and 4.4~eV. Given the
PDOS in Fig.~\ref{fig:bands-hse}, these peaks are consistent with transitions from Te-5p
valence states to the conduction band. Above $\approx$4.4~eV, $\varepsilon_2(\omega)$
decreases because the joint density of states for interband transitions falls.

The real part of the optical conductivity (Fig.~\ref{fig:optical1}(c)) is proportional to
$\omega\varepsilon_2(\omega)$ according to Eq.~(\ref{eq:sigma}). It rises from the
absorption onset near 1~eV and peaks at $\approx8\times10^{15}$~s$^{-1}$ near 4.5~eV,
which is consistent with $\varepsilon_2 \approx 13$ at this energy. This quantity
describes interband absorption; it is distinct from the dc electrical conductivity, which
depends on free carriers and is treated in Section~\ref{sec:thermo}.

Fig.~\ref{fig:optical1}(d) shows the absorption coefficient. $\alpha(\omega)$ rises
steeply above the onset, reaching $\approx3.5\times10^{5}$~cm$^{-1}$ at 2.3~eV, and attains
its maximum of $\approx1.45\times10^{6}$~cm$^{-1}$ at 6.5--7.5~eV. The corresponding
absorption length $1/\alpha$ is below 30~nm at 2.3~eV, so visible light above the gap is
absorbed within a few hundred nanometers. ZnGa$_2$Te$_4$ absorbs slightly more strongly
than CdGa$_2$Te$_4$ over much of the visible and near-ultraviolet range. Because the PBEsol
onset lies $\approx$1.2~eV (CdGa$_2$Te$_4$) and $\approx$0.9~eV (ZnGa$_2$Te$_4$) below the
HSE06 gaps, the PBEsol $\alpha(\omega)$ predicts absorption in an energy range where the
HSE06 band structure has no allowed transitions. This must be considered when the optical
constants are transferred to device simulations (Section~\ref{sec:fdtd-method}).

Fig.~\ref{fig:optical2}(a) shows the reflectivity. $R(0)$ is $\approx$0.28 for both
compounds, consistent with Eq.~(\ref{eq:R}) and the static refractive index. $R(\omega)$
rises to $\approx$0.40 near 2~eV and reaches its highest values of $\approx$0.50--0.55
between $\approx$7 and 11~eV, where $\varepsilon_1(\omega)$ is negative. The appreciable
reflectivity in the visible range implies that front-surface anti-reflection measures are
needed in devices, as confirmed by the reflection losses in Section~\ref{sec:absorption}.

Figs.~\ref{fig:optical2}(b) and \ref{fig:optical2}(c) show $k(\omega)$ and $n(\omega)$.
$k(\omega)$ rises from the absorption onset to $\approx$1.5 at 2.3~eV, in agreement with
the value of $\alpha$ given above through Eq.~(\ref{eq:alpha}), and reaches its maximum of
$\approx$2.5--2.7 near 4.8~eV before decreasing. The static refractive index $n(0)$ is
$\approx$3.25--3.3, consistent with $\sqrt{\varepsilon_1(0)}$. $n(\omega)$ increases to a
maximum of $\approx$4.2 near 2~eV, as expected for normal dispersion below the main
absorption band, and decreases at higher energies as $\varepsilon_1(\omega)$ falls.
Fig.~\ref{fig:optical2}(d) shows the energy-loss function $L(\omega)$, which describes the
energy lost by fast electrons traversing the material. Its main peak, at
$\approx$12.2~eV for CdGa$_2$Te$_4$ and $\approx$12.5~eV for ZnGa$_2$Te$_4$, marks the bulk
plasma energy, at which collective oscillations of the valence electrons are excited
\cite{segall2002,naher2022}; it coincides with the sharp drop of the reflectivity at the
high-energy end of Fig.~\ref{fig:optical2}(a). The behavior of the photogenerated
electron--hole pairs is examined next through their excitonic properties.

\subsection{Excitonic properties}
\label{sec:excitonic}

\subsubsection{Exciton binding energy}
\label{sec:Eb}

Photogenerated electrons and holes attract each other and can form excitons. If the
exciton binding energy $E_b$ is large compared with $k_{\mathrm{B}}T$, excitons must be
separated at heterointerfaces, as in organic solar cells \cite{liu2018}. If $E_b$ is comparable to or
smaller than $k_{\mathrm{B}}T$, they dissociate thermally into free carriers, as required
for conventional p--n junction cells. Within the Wannier--Mott model, which applies when
the exciton extends over many unit cells, $E_b$ is estimated from
\cite{wang2017,wannier1937}
\begin{equation}
  E_b = \frac{13.6~\mathrm{eV}\,\left(\mu^*/m_0\right)}{\varepsilon_s^{2}} ,
  \qquad
  \mu^* = \frac{m_e^*\, m_h^*}{m_e^* + m_h^*} ,
  \label{eq:Eb}
\end{equation}
where $\mu^*$ is the reduced effective mass, $m_0$ is the free-electron mass and
$\varepsilon_s$ is the static dielectric constant. With the effective masses in
Table~\ref{tab:mass} ($\mu^* = 0.173\,m_0$ and $0.267\,m_0$), Eq.~(\ref{eq:Eb}) gives
$E_b = 18.85$~meV for CdGa$_2$Te$_4$ and 26.81~meV for ZnGa$_2$Te$_4$, corresponding to
$\varepsilon_s \approx 11.2$ and $\approx 11.6$, respectively, taken from the calculated zero-frequency dielectric constant $\varepsilon_1(0)$ (Section~\ref{sec:optical}).
Both binding energies are comparable to the thermal energy $k_{\mathrm{B}}T = 25.9$~meV at 300~K and lie
within the range reported for MAPbI$_3$ (2--75~meV) \cite{umari2018}.

The spatial extent of the exciton is described by the exciton Bohr radius $R_{ex}$, and
its thermal stability by the dissociation temperature $T_{ex}$, at which $k_{\mathrm{B}}T$
equals $E_b$ \cite{wannier1937,hirasawa1994}:
\begin{align}
  R_{ex} &= \varepsilon_s\,\frac{m_0}{\mu^*}\, n^{2}\, a_{\mathrm{B}} ,
  \label{eq:Rex} \\
  T_{ex} &= \frac{E_b}{k_{\mathrm{B}}} ,
  \label{eq:Tex}
\end{align}
where $n$ is the principal quantum number ($n = 1$ for the ground state) and
$a_{\mathrm{B}} = 0.529$~\AA{} is the hydrogen Bohr radius. The calculated $R_{ex}$ values
of 34.3~\AA{} (CdGa$_2$Te$_4$) and 23.0~\AA{} (ZnGa$_2$Te$_4$) are about five and four
times the lattice constant $a$. The excitons are therefore delocalized over several unit
cells, which justifies the Wannier--Mott description. The dissociation temperatures are
218.7 and 311.1~K. At 300~K, excitons in CdGa$_2$Te$_4$ are thermally unstable, whereas
those in ZnGa$_2$Te$_4$ are marginally stable, so a fraction may persist as bound pairs.
In both cases, the built-in field of the heterojunction (Section~\ref{sec:alignment}) is
expected to assist their dissociation. CdGa$_2$Te$_4$, with the lower $E_b$ and larger
$R_{ex}$, offers slightly more favorable conditions for free-carrier generation.

\subsection{Thermoelectric properties}
\label{sec:thermo}

\begin{figure*}[pos=t]
  \centering
  \includegraphics[width=0.7\textwidth]{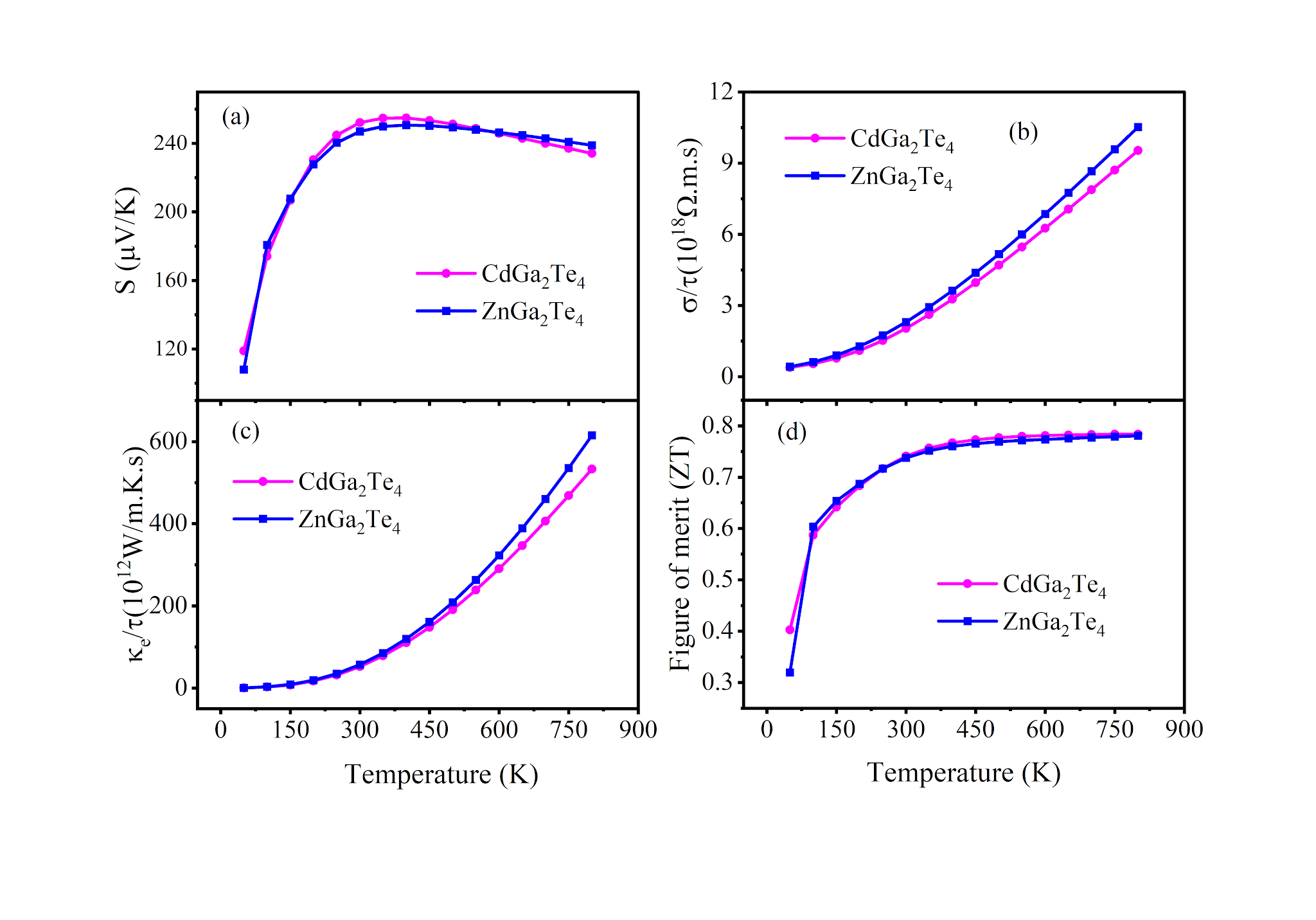}
  \caption{Temperature dependence of (a) the Seebeck coefficient $S$, (b) the electrical
           conductivity per relaxation time $\sigma/\tau$, (c) the electronic thermal
           conductivity per relaxation time $\kappa_e/\tau$ and (d) the electronic figure
           of merit $ZT_e$ of CdGa$_2$Te$_4$ and ZnGa$_2$Te$_4$, calculated in the constant
           relaxation time approximation at a fixed chemical potential corresponding to intrinsic carrier conditions at 300~K.}
  \label{fig:thermo-T}
\end{figure*}

The thermoelectric transport coefficients were calculated from the PBEsol band structure
with semiclassical Boltzmann transport theory in the CRTA (Section~\ref{sec:dft-method}).
The Seebeck coefficient is the open-circuit thermovoltage per unit temperature difference,
\begin{equation}
  S = \frac{\Delta V}{\Delta T} .
  \label{eq:seebeck}
\end{equation}
In the CRTA, $S$ is independent of the relaxation time $\tau$, whereas the electrical
conductivity $\sigma$ and the electronic thermal conductivity $\kappa_e$ are obtained only
as $\sigma/\tau$ and $\kappa_e/\tau$ \cite{madsen2006,snyder2008}. Fig.~\ref{fig:thermo-T}
shows these quantities between 50 and 800~K. The thermoelectric figure of merit and its
electronic part are defined as
\begin{align}
  ZT &= \frac{S^{2}\sigma T}{\kappa_e + \kappa_l} ,
  \label{eq:zt} \\
  ZT_e &= \frac{S^{2}\sigma T}{\kappa_e}
        = \frac{S^{2}(\sigma/\tau)\,T}{\kappa_e/\tau} ,
  \label{eq:zte}
\end{align}
where $\kappa_l$ is the lattice thermal conductivity. Because $\tau$ cancels in
Eq.~(\ref{eq:zte}), $ZT_e$ can be evaluated within the CRTA without knowledge of $\tau$.
$\kappa_l$ was not calculated in this work, so $ZT_e$ is an upper bound on $ZT$ that is
approached only when $\kappa_l \ll \kappa_e$.

The Seebeck coefficient (Fig.~\ref{fig:thermo-T}(a)) is positive over the whole
temperature range, which indicates hole-dominated transport under the conditions of the
calculation. $S$ increases from $\approx$110--120~$\mu$V~K$^{-1}$ at 50~K to a broad
maximum of $\approx$255~$\mu$V~K$^{-1}$ (CdGa$_2$Te$_4$) and $\approx$250~$\mu$V~K$^{-1}$
(ZnGa$_2$Te$_4$) at 350--400~K, and then decreases slightly to
$\approx$234--239~$\mu$V~K$^{-1}$ at 800~K. In a semiconductor, a maximum in $S(T)$ is
often attributed to the onset of bipolar conduction, and the Goldsmid--Sharp relation
$E_g \approx 2eS_{\max}T_{\max}$ then estimates the gap \cite{goldsmid1999}. Here this
relation gives only $\approx$0.2~eV, far below both the PBEsol and HSE06 gaps. The maximum
is therefore unlikely to arise from minority-carrier excitation across the gap; it more
probably reflects the temperature dependence of the carrier distribution at the fixed
chemical potential, which was held fixed. The
two compounds have almost identical $S(T)$, consistent with their similar Te-5p-dominated
valence bands.

$\sigma/\tau$ (Fig.~\ref{fig:thermo-T}(b)) increases monotonically, from
$\approx0.4\times10^{18}$~$\Omega^{-1}$~m$^{-1}$~s$^{-1}$ at 50~K to
$\approx2\times10^{18}$~$\Omega^{-1}$~m$^{-1}$~s$^{-1}$ at 300~K, and to
$\approx9.5\times10^{18}$ (CdGa$_2$Te$_4$) and
$\approx10.5\times10^{18}$~$\Omega^{-1}$~m$^{-1}$~s$^{-1}$ (ZnGa$_2$Te$_4$) at 800~K,
reflecting the growing number of thermally excited carriers within the transport window.
$\kappa_e/\tau$ (Fig.~\ref{fig:thermo-T}(c)) follows the same trend, rising from
$\approx52\times10^{12}$~W~m$^{-1}$~K$^{-1}$~s$^{-1}$ at 300~K to $\approx535\times10^{12}$
(CdGa$_2$Te$_4$) and $\approx615\times10^{12}$~W~m$^{-1}$~K$^{-1}$~s$^{-1}$
(ZnGa$_2$Te$_4$) at 800~K. ZnGa$_2$Te$_4$ has slightly higher $\sigma/\tau$ and
$\kappa_e/\tau$ at high temperature. Absolute values of $\sigma$ and $\kappa_e$ require
$\tau$, which was not determined here; for an assumed $\tau = 10^{-14}$~s, $\sigma$ at
300~K would be $\approx2\times10^{4}$~$\Omega^{-1}$~m$^{-1}$, so the absolute
conductivities are indicative only.

The ratio of the two conductivities defines an apparent Lorenz number,
\begin{equation}
  L = \frac{\kappa_e}{\sigma T} ,
  \label{eq:lorenz}
\end{equation}
which is independent of $\tau$. From Figs.~\ref{fig:thermo-T}(b, c),
$L \approx 8.7\times10^{-8}$~W~$\Omega$~K$^{-2}$ at 300~K and
$\approx$7.0--7.3$\times10^{-8}$~W~$\Omega$~K$^{-2}$ at 800~K. These values are about three
times the degenerate (Sommerfeld) limit of $2.44\times10^{-8}$~W~$\Omega$~K$^{-2}$ and four
to five times the value of $\approx1.6\times10^{-8}$~W~$\Omega$~K$^{-2}$ expected for a
non-degenerate semiconductor with $|S| \approx$~235--250~$\mu$V~K$^{-1}$ \cite{kim2015}.
Such a large $L$ would normally indicate a strong bipolar contribution, which is unlikely
given the Goldsmid--Sharp analysis above. A more plausible explanation is that
Fig.~\ref{fig:thermo-T}(c) shows the thermal conductivity at zero electric field,
$\kappa_0$, which exceeds the zero-current electronic thermal conductivity by
$T\sigma S^2$:
\begin{equation}
  \kappa_e = \kappa_0 - T\sigma S^{2} .
  \label{eq:kappa0}
\end{equation}
Applying Eq.~(\ref{eq:kappa0}) at 800~K gives $L \approx 1.6\times10^{-8}$~W~$\Omega$~K$^{-2}$
for both compounds, in agreement with the non-degenerate value. Consistently, the $ZT_e$
values in Fig.~\ref{fig:thermo-T}(d) are reproduced by $S^2\sigma T$ divided by the
plotted conductivity, consistent with the quantity shown in Fig.~\ref{fig:thermo-T}(c) being $\kappa_0$ and $ZT_e$ being evaluated from $\kappa_e$ via Eq.~(\ref{eq:kappa0}) and Eq.~(\ref{eq:zte}).

As plotted, $ZT_e$ (Fig.~\ref{fig:thermo-T}(d)) increases from $\approx$0.40
(CdGa$_2$Te$_4$) and $\approx$0.32 (ZnGa$_2$Te$_4$) at 50~K to $\approx$0.74 at 300~K and
approaches $\approx$0.78 at 800~K. Above $\approx$350~K, the two compounds differ only
marginally, with CdGa$_2$Te$_4$ slightly higher. Because $\kappa_e/\tau$ and $\sigma/\tau$
increase almost in proportion, $ZT_e$ is controlled mainly by $S$ and saturates as $S$
levels off. These values are upper limits with respect to $\kappa_l$, and $ZT$ will be
substantially lower wherever $\kappa_l$ is comparable to $\kappa_e$. For comparison,
Govindaraj \emph{et al.} obtained a full figure of merit of 0.77 at 750~K for
ZnGa$_2$Te$_4$ only under 12~GPa, where $\kappa_l$ drops to $\approx$1~W~m$^{-1}$~K$^{-1}$
\cite{govindaraj2022}, which shows how strongly $\kappa_l$ controls the attainable $ZT$.

The lattice contribution can be assessed only qualitatively. The phonon spectra
(Section~\ref{sec:structure}) extend only to $\approx$6.5~THz, with acoustic branches below
$\approx$2~THz and low-frequency modes dominated by heavy Te atoms. Such features generally
imply low phonon group velocities and a low $\kappa_l$ \cite{snyder2008}, but a
quantitative value requires anharmonic phonon calculations. Absolute $ZT$ values also
require $\tau$, which can be estimated from deformation-potential theory or explicit
electron--phonon calculations. Within these limits, both compounds combine a large
Seebeck coefficient with a promising electronic figure of merit and merit further
thermoelectric study. The thermal conductivity discussed here is a material transport
coefficient and should not be confused with the optical heat-source density of
Section~\ref{sec:genheat}, which is a volumetric heat-generation rate; for a given heat
source, the temperature rise in a device scales inversely with the total thermal
conductivity $\kappa_e + \kappa_l$.

\subsection{Photocatalytic properties}
\label{sec:photocatalytic}

\begin{figure}[pos=t]
    \centering
    \includegraphics[width=0.95\columnwidth]{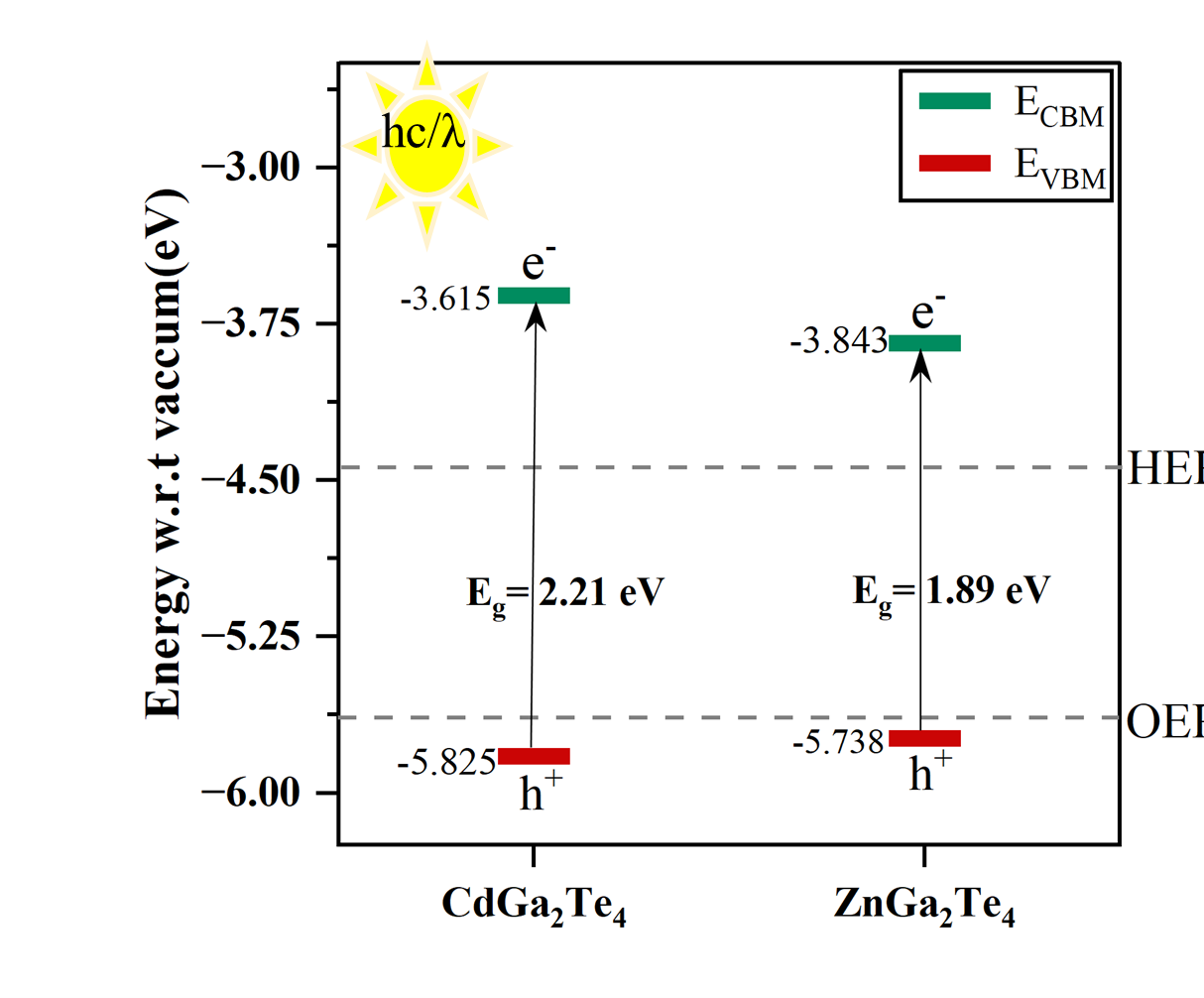}
    \caption{Band-edge positions of CdGa$_2$Te$_4$ and ZnGa$_2$Te$_4$ relative to the vacuum
    level, estimated from the Mulliken electronegativity and the HSE06 gaps,
    compared with the H$^+$/H$_2$ and O$_2$/H$_2$O redox potentials at pH~0.}
    \label{fig:water}
\end{figure}

Photocatalytic water splitting converts solar energy into hydrogen without carbon
emissions \cite{kudo2009}. The overall reaction is
\begin{equation}
  2\,\text{H}_2\text{O} + \text{catalyst} \rightarrow 2\,\text{H}_2 + \text{O}_2
  + \text{catalyst} .
  \label{eq:water}
\end{equation}
A single semiconductor can drive this reaction only if its band gap exceeds the 1.23~eV
thermodynamic requirement plus the kinetic overpotentials, and if its band edges straddle
the water redox potentials, with the CBM above the H$^+$/H$_2$ level and the VBM below the
O$_2$/H$_2$O level \cite{kudo2009}. The HSE06 gaps of 2.21~eV (CdGa$_2$Te$_4$) and
1.895~eV (ZnGa$_2$Te$_4$) satisfy the first criterion and correspond to absorption edges at
$\approx$561 and $\approx$654~nm in the visible range. On the absolute (vacuum) scale at
pH~0, the H$^+$/H$_2$ and O$_2$/H$_2$O potentials lie at $-4.44$ and $-5.67$~eV,
respectively. The band-edge energies relative to the vacuum level were estimated with the
empirical Mulliken electronegativity approach \cite{butler1978},
\begin{align}
  E_{\mathrm{CBM}} &= -\left(\chi - \frac{E_g}{2}\right) ,
  \label{eq:ecbm} \\
  E_{\mathrm{VBM}} &= -\left(\chi + \frac{E_g}{2}\right) ,
  \label{eq:evbm}
\end{align}
where $\chi$ is the Mulliken electronegativity of the compound, taken as the geometric
mean of the atomic electronegativities \cite{butler1978},
and $E_g$ is the HSE06 gap. $\chi$ is 4.72~eV for CdGa$_2$Te$_4$ and 4.79~eV for
ZnGa$_2$Te$_4$.

The resulting CBM and VBM energies are $-3.615$ and $-5.825$~eV for CdGa$_2$Te$_4$ and
$-3.843$ and $-5.738$~eV for ZnGa$_2$Te$_4$ (Fig.~\ref{fig:water}). In both compounds,
the CBM lies above the H$^+$/H$_2$ level and the VBM lies below the O$_2$/H$_2$O level, so
the band edges straddle both redox potentials at pH~0. The driving force for the hydrogen
evolution reaction (HER), defined as the offset between the CBM and the H$^+$/H$_2$ level,
is 0.825~eV for CdGa$_2$Te$_4$ and 0.597~eV for ZnGa$_2$Te$_4$. The driving force for the
oxygen evolution reaction (OER), defined as the offset between the VBM and the
O$_2$/H$_2$O level, is much smaller: 0.155~eV for CdGa$_2$Te$_4$ and 0.068~eV for
ZnGa$_2$Te$_4$. These margins are below the overpotentials usually required for water
oxidation, so the OER is expected to be kinetically limiting, particularly for
ZnGa$_2$Te$_4$ \cite{kudo2009}. Both compounds are therefore better suited to hydrogen
evolution, for example with a sacrificial hole scavenger or in a Z-scheme, than to
unassisted overall water splitting. The Mulliken approach also neglects surface
termination, surface dipoles and pH-dependent band bending, so the small OER margins in
particular should be confirmed by slab calculations or flat-band potential measurements.
Like other metal chalcogenides, these tellurides may also be susceptible to oxidative
photocorrosion in aqueous media \cite{kudo2009}, which must be assessed before
photocatalytic use.

\subsection{Photovoltaic device analysis of CdGa$_2$Te$_4$ and ZnGa$_2$Te$_4$ absorbers}
\label{sec:pv}

The first-principles results identify CdGa$_2$Te$_4$ and ZnGa$_2$Te$_4$ as direct-gap
semiconductors with strong optical absorption and weakly bound excitons. Whether these
properties translate into efficient light-to-electricity conversion depends on how each
absorber interacts with the contacts, the optical field and the carrier-transport pathways
of a complete device. Both compounds were therefore examined in the
FTO/IGZO/XGa$_2$Te$_4$/CuAlO$_2$/Pt architecture of Section~\ref{sec:device-method}, using
the coupled FDTD--CHARGE framework. This section analyzes, in turn, the interfacial band
alignment, the optical absorption and photogeneration obtained from FDTD, and the
electrical response obtained from CHARGE, and closes with a comparative assessment of the
two absorbers.

\subsubsection{Energy-band alignment}
\label{sec:alignment}

The simulated energy-band diagrams in Fig.~\ref{fig:device}(c, d) define the electrostatic
landscape that governs carrier separation in both devices. In each absorber, the
conduction- and valence-band edges remain flat between approximately $-0.6$ and
0.45~$\mu$m, marking a quasi-neutral region, whereas strong band bending is confined to
the front $\approx$0.3~$\mu$m adjacent to IGZO. This front space-charge region coincides
with the zone of highest photogeneration (Section~\ref{sec:genheat}), so most
photogenerated electrons are created where the built-in field can sweep them towards IGZO.
A narrower region of band bending, $\approx$0.1--0.15~$\mu$m wide, also forms at the rear
interface with CuAlO$_2$.

At the front interface, the conduction-band offset ($\Delta E_{\mathrm{C}}$) is +0.15~eV
for CdGa$_2$Te$_4$ and +0.33~eV for ZnGa$_2$Te$_4$. A positive value here denotes an IGZO
conduction band lying above that of the absorber, i.e.\ a spike-type alignment. Device
simulations and experiments on planar heterojunctions show that a small spike suppresses
interfacial recombination, whereas a cliff-type offset makes interface recombination
dominant and lowers $V_{\mathrm{OC}}$ \cite{wang2023,minemoto2015,ding2018}. The optimum
spike has been placed at 0--0.3~eV, beyond which the barrier begins to impede electron
collection \cite{minemoto2015}. The 0.15~eV offset of CdGa$_2$Te$_4$ lies within this
window, whereas the 0.33~eV offset of ZnGa$_2$Te$_4$ lies marginally above it. The large
valence-band offsets at the same interface, $-0.69$~eV (CdGa$_2$Te$_4$) and $-0.91$~eV
(ZnGa$_2$Te$_4$), confirm that IGZO acts as an effective hole-blocking layer, consistent
with its wide band gap and n-type character \cite{nomura2004}.

At the rear interface, the conduction-band offsets of +1.84~eV (CdGa$_2$Te$_4$) and
+2.01~eV (ZnGa$_2$Te$_4$) create a high barrier that reflects minority electrons away from
the hole contact, a role analogous to a back-surface field \cite{biplab2020bsf,gezgin2026}.
Hole extraction into CuAlO$_2$, a wide-gap p-type transparent oxide \cite{kawazoe1997}, is
less ideal. The CuAlO$_2$ valence band lies 0.59~eV (CdGa$_2$Te$_4$) and 0.36~eV
(ZnGa$_2$Te$_4$) above that of the absorber, a valence-band cliff that favors interfacial
recombination \cite{minemoto2015}. Because of this cliff, the equilibrium Fermi level of
the heavily doped p-type CuAlO$_2$ lies above that of the moderately doped absorber.
Fermi-level equilibration therefore transfers electrons into the absorber side of the
rear interface, bending both bands downwards there, depleting holes and producing a small
barrier to hole transport (Fig.~\ref{fig:device}(c, d)). The barrier is larger for the
larger cliff and has a visible consequence in the CdGa$_2$Te$_4$ device:
$E_{\mathrm{Fp}}$ drops by $\approx$0.18~eV across the CuAlO$_2$/absorber interface,
whereas it remains continuous in the ZnGa$_2$Te$_4$ device. A quasi-Fermi-level step of
this kind signals a loss in hole extraction at the rear contact, which is examined further
through the fill factor in Section~\ref{sec:jv}. Overall, the band diagrams show that the
front junction is well aligned in both devices, and that the rear contact is the more
critical interface for CdGa$_2$Te$_4$.

\subsubsection{Optical absorption and thin-film interference}
\label{sec:absorption}

\begin{figure*}[pos=t]
  \centering
  \includegraphics[width=0.9\textwidth]{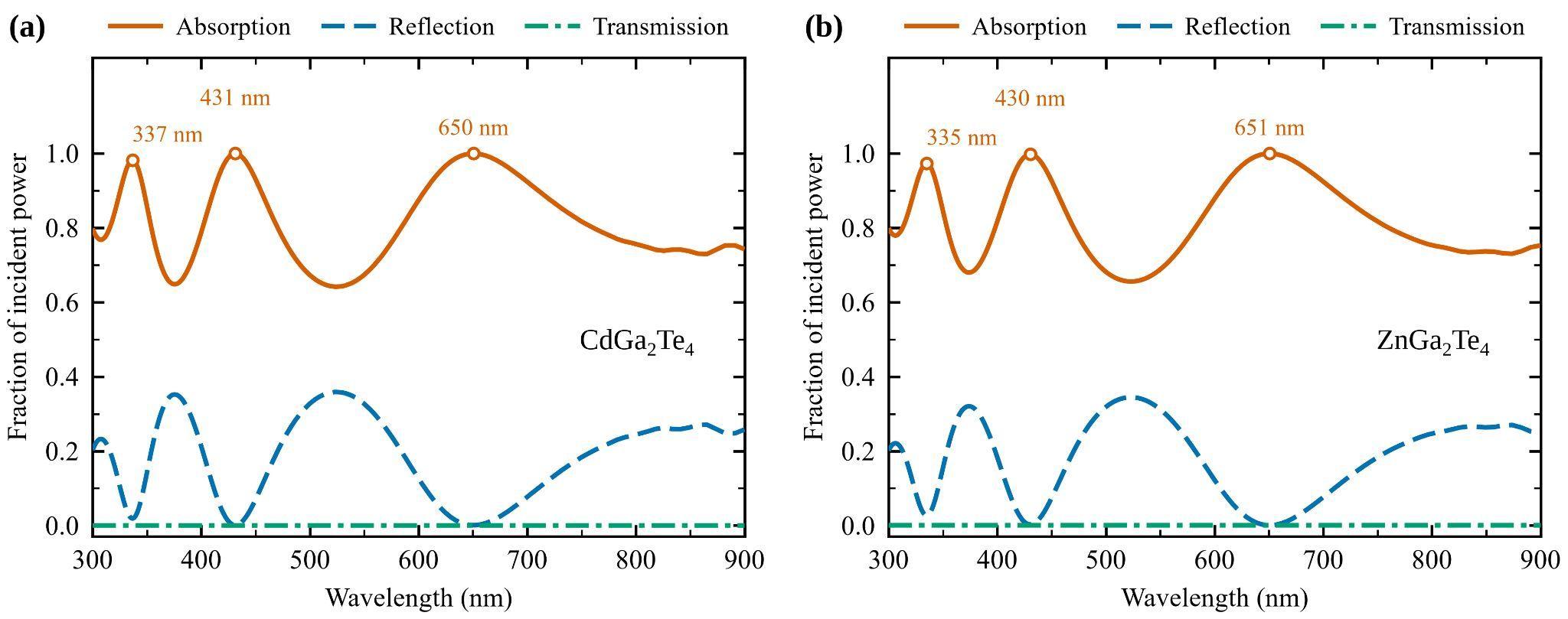}
  \caption{FDTD-simulated wavelength-dependent absorption $A(\lambda)$, reflection
           $R(\lambda)$ and transmission $T(\lambda)$ of the (a) CdGa$_2$Te$_4$- and
           (b) ZnGa$_2$Te$_4$-based device stacks between 300 and 900~nm.}
  \label{fig:ART}
\end{figure*}

\FloatBarrier

The optical response was analyzed first, because it sets the upper bound on
photogeneration that electrical transport can only reduce. Fig.~\ref{fig:ART} shows
$A(\lambda)$, $R(\lambda)$ and $T(\lambda)$ of the two device stacks between 300 and
900~nm. $T(\lambda)$ is essentially zero across the whole range, confirming that the
100~nm Pt electrode acts as an opaque back reflector; consequently,
$A(\lambda) = 1 - R(\lambda)$ according to Eq.~(\ref{eq:ART}), and reflection is the only
optical loss channel at the stack level. The absorption spectra of the CdGa$_2$Te$_4$ and
ZnGa$_2$Te$_4$ devices are almost identical. Both show three maxima of $\approx$0.98--1.00,
at 337, 431 and 650~nm for CdGa$_2$Te$_4$ and at 335, 430 and 651~nm for ZnGa$_2$Te$_4$.
Between these maxima, absorption falls to $\approx$0.64--0.68 near 375 and 525~nm, exactly
where $R(\lambda)$ peaks at $\approx$0.32--0.36. Above $\approx$700~nm, $A(\lambda)$ levels
off at $\approx$0.73--0.80 while $R(\lambda)$ rises towards $\approx$0.27. Absorption at
wavelengths longer than $\approx$561~nm (CdGa$_2$Te$_4$) and $\approx$654~nm
(ZnGa$_2$Te$_4$), that is, below the HSE06 gaps, must originate either from the PBEsol-based
absorber optical constants (Section~\ref{sec:fdtd-method}) or from parasitic absorption in
the electrodes. Only the former contributes to the simulated photogeneration in the
absorber, and it would vanish with gap-corrected optical constants.

The near coincidence of the spectral features for two different absorbers shows that the
modulation is governed not by the absorbers themselves but by thin-film interference in
the 200~nm FTO and 50~nm IGZO front stack \cite{pettersson1999,burkhard2010}. This
assignment is supported by the field distributions in Section~\ref{sec:thickness}: at 350
and 552~nm, light is extinguished within $\approx$0.1--0.15~$\mu$m of the absorber
surface, so cavity resonances across the 1.5~$\mu$m absorber cannot shape the
short-wavelength spectrum. For a single interference layer of thickness $d$ and mean
refractive index $\bar{n}$ on a higher-index substrate, reflection minima occur at
\begin{equation}
  \lambda_m = \frac{4\bar{n}d}{2m+1}, \qquad m = 0, 1, 2, \ldots
  \label{eq:fringe}
\end{equation}
Taking $d = 250$~nm for the combined FTO/IGZO stack, the absorption maxima at 650, 431
and 337~nm correspond to $m = 1$, 2 and 3 with $\bar{n} \approx 1.95$, 2.16 and 2.36,
respectively. The rise of $\bar{n}$ towards shorter wavelengths is consistent with normal
dispersion in the oxide layers, which supports the interpretation of the front stack as an
interference coating. The practical implication is that the $\approx$30--35\% reflection
at the spectral minima is the largest recoverable optical loss in both devices
\cite{oni2024}. Tuning the FTO or IGZO thickness, or adding a broadband anti-reflection
layer, would shift and flatten these fringes without changing the absorber.

\subsubsection{Two-dimensional photogeneration and heat-source distributions}
\label{sec:genheat}

\begin{figure*}[pos=t]
  \centering
  \includegraphics[width=0.8\textwidth]{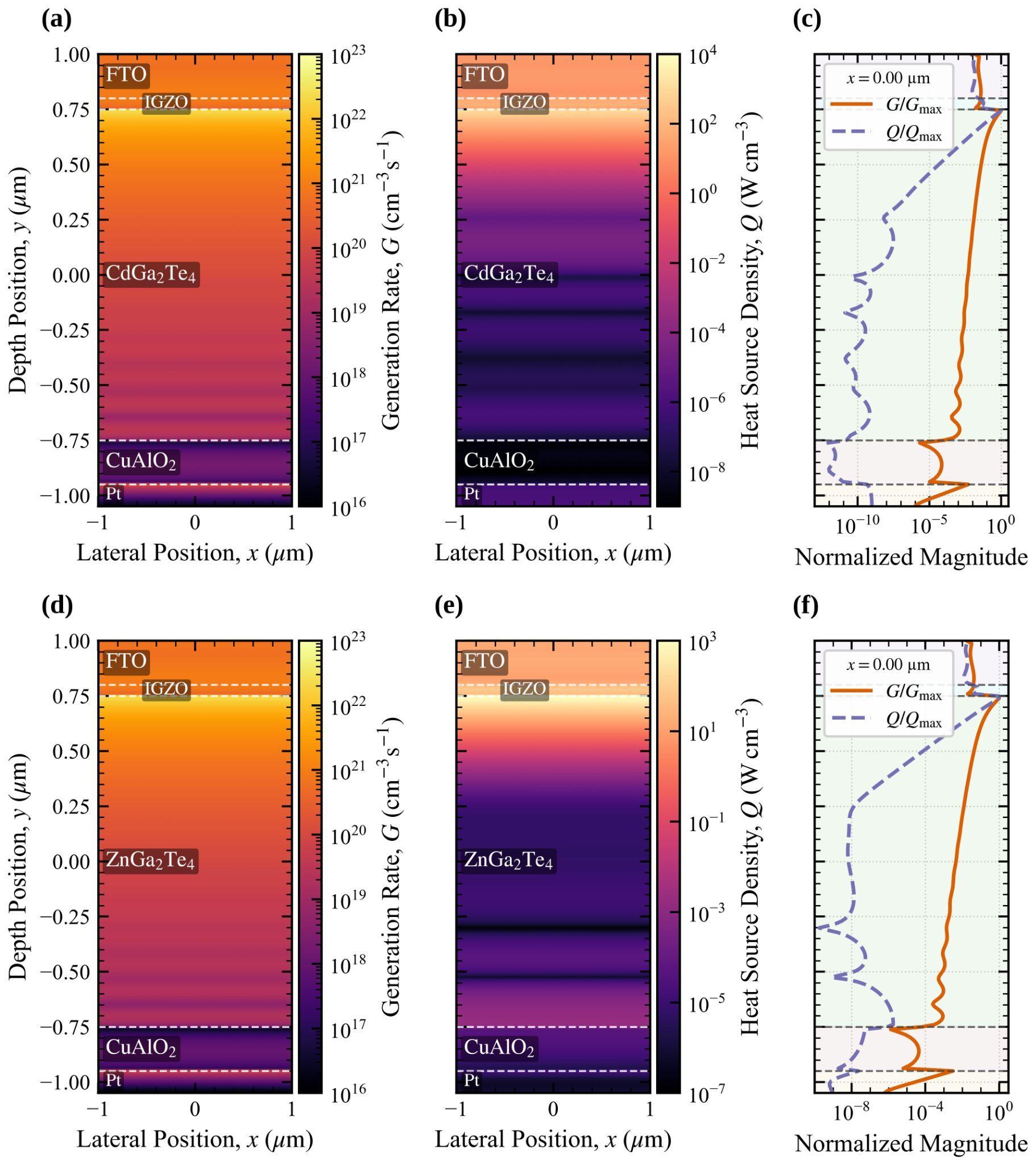}
  \caption{FDTD-derived spatial distributions of (a, d) the optical generation rate $G$
           (cm$^{-3}$~s$^{-1}$, Eq.~(\ref{eq:gen})), (b, e) the heat-source density $Q$
           (W~cm$^{-3}$), and (c, f) the corresponding normalized depth profiles
           $G/G_{\max}$ and $Q/Q_{\max}$ at $x = 0$ in solar cells with (a--c)
           CdGa$_2$Te$_4$ and (d--f) ZnGa$_2$Te$_4$ absorbers. All color scales are
           logarithmic.}
  \label{fig:genheat}
\end{figure*}

\FloatBarrier

The 2D maps in Fig.~\ref{fig:genheat} resolve where absorbed photons are converted into
carriers and where absorbed power is deposited as heat. Both $G$ and $Q$ are laterally
uniform, as expected for a planar stack without lateral structure, so the depth profiles
at $x = 0$ (Fig.~\ref{fig:genheat}(c, f)) represent the entire device. In both devices,
$G$ peaks at the IGZO/absorber interface ($y \approx 0.75$~$\mu$m), where it reaches the
$10^{22}$--$10^{23}$~cm$^{-3}$~s$^{-1}$ range, and decreases by roughly two to three orders
of magnitude towards the rear of the absorber \cite{wahba2023,roy2024}. In the rear half
of the absorber, $G$ shows a periodic modulation that reflects standing waves of weakly
absorbed long-wavelength light returned by the Pt mirror. Generation in CuAlO$_2$ and Pt
is about four orders of magnitude below the peak, so parasitic absorption at the back of
the device is negligible. By contrast, the FTO layer shows a generation level only about
two orders of magnitude below the absorber peak. Because FTO is modeled as a conductor,
absorption there cannot contribute to the photocurrent; it is a parasitic loss that is
included in $A(\lambda)$ but not in the collectable generation \cite{burkhard2010}.

The depth distribution of $G$ must be compared with the distance over which carriers can
be collected. The minority-carrier diffusion length follows from
\begin{equation}
  L_{n,p} = \sqrt{D_{n,p}\tau_{n,p}}
          = \sqrt{\frac{k_{\mathrm{B}}T}{q}\,\mu_{n,p}\,\tau_{n,p}} .
  \label{eq:diffL}
\end{equation}
With the parameters in Table~\ref{tab:charge} and $T = 300$~K, Eq.~(\ref{eq:diffL}) gives
electron diffusion lengths of $\approx$198~nm (CdGa$_2$Te$_4$) and $\approx$213~nm
(ZnGa$_2$Te$_4$); with the 1~ns hole lifetime, the hole diffusion lengths are
$\approx$171 and $\approx$214~nm. Electrons, the minority carriers in the p-type
absorbers, are therefore collected efficiently only within approximately the front
space-charge width plus one diffusion length, that is, within $\approx$0.5~$\mu$m of the
IGZO interface. The front-heavy generation profile places most photogenerated electrons
inside this collection zone, so a large fraction of them can be collected despite the
short electron lifetime.

The heat-source density $Q$ (Fig.~\ref{fig:genheat}(b, e)) is concentrated in the FTO
layer and in the first $\approx$0.2~$\mu$m of the absorber, with color-scale maxima of
$\approx10^{4}$~W~cm$^{-3}$ (CdGa$_2$Te$_4$) and $\approx10^{3}$~W~cm$^{-3}$
(ZnGa$_2$Te$_4$). Through the absorber, $Q$ decreases by $\approx$8--10 orders of
magnitude, whereas $G$ decreases by only $\approx$2--3 orders
(Fig.~\ref{fig:genheat}(c, f)). The definitions in Section~\ref{sec:gq-method} allow these
observations to be tested. If $Q$ were the total absorbed power density $Q_{\mathrm{abs}}$
of Eq.~(\ref{eq:qabs}), the ratio $Q/(qG)$ would equal the mean energy of the absorbed
photons and would lie between 1.38 and 4.13~eV at every depth; $Q$ and $G$ would then
decay in near proportion. Order-of-magnitude estimates from the color scales give
$Q/(qG)$ of order 1~eV near the front of the CdGa$_2$Te$_4$ absorber, close to this range,
but of order 0.1~eV already at the front of ZnGa$_2$Te$_4$, and far smaller values deeper
in both absorbers, which rules out $Q = Q_{\mathrm{abs}}$. If instead $Q$ were the
thermalization heat $Q_{\mathrm{th}}$ of Eq.~(\ref{eq:qth}), it would indeed decay faster
than $G$, because the photons surviving deep into the absorber have energies close to the
absorption edge and carry little excess energy; however, $Q_{\mathrm{th}}$ would then be
larger for the narrower-gap ZnGa$_2$Te$_4$, contrary to Fig.~\ref{fig:genheat}. Neither
definition is therefore consistent with both observations, and the heat-source maps are
reported here descriptively. What can be stated
robustly is that both carrier generation and heat deposition are front-loaded: the
heat load is concentrated in the FTO electrode and the junction region, which is where
thermally activated degradation would be expected first \cite{dupre2016}.

\subsubsection{Effect of absorber doping density}
\label{sec:doping}

The optical analysis establishes how many carriers are generated and where. The electrical
analysis determines how many of them are extracted, and how the absorber doping and
thickness control that fraction. Because the imported generation rate includes absorption
below the HSE06 gaps (Section~\ref{sec:absorption}), the absolute $J_{\mathrm{SC}}$ and PCE
values in this and the following sections are overestimated (Section~\ref{sec:jv}). The
trends with doping and thickness, which are governed mainly by the space-charge width,
the diffusion length and optical interference, are discussed as relative design guidance.

\begin{figure*}[pos=t]
  \centering
  \includegraphics[width=\textwidth]{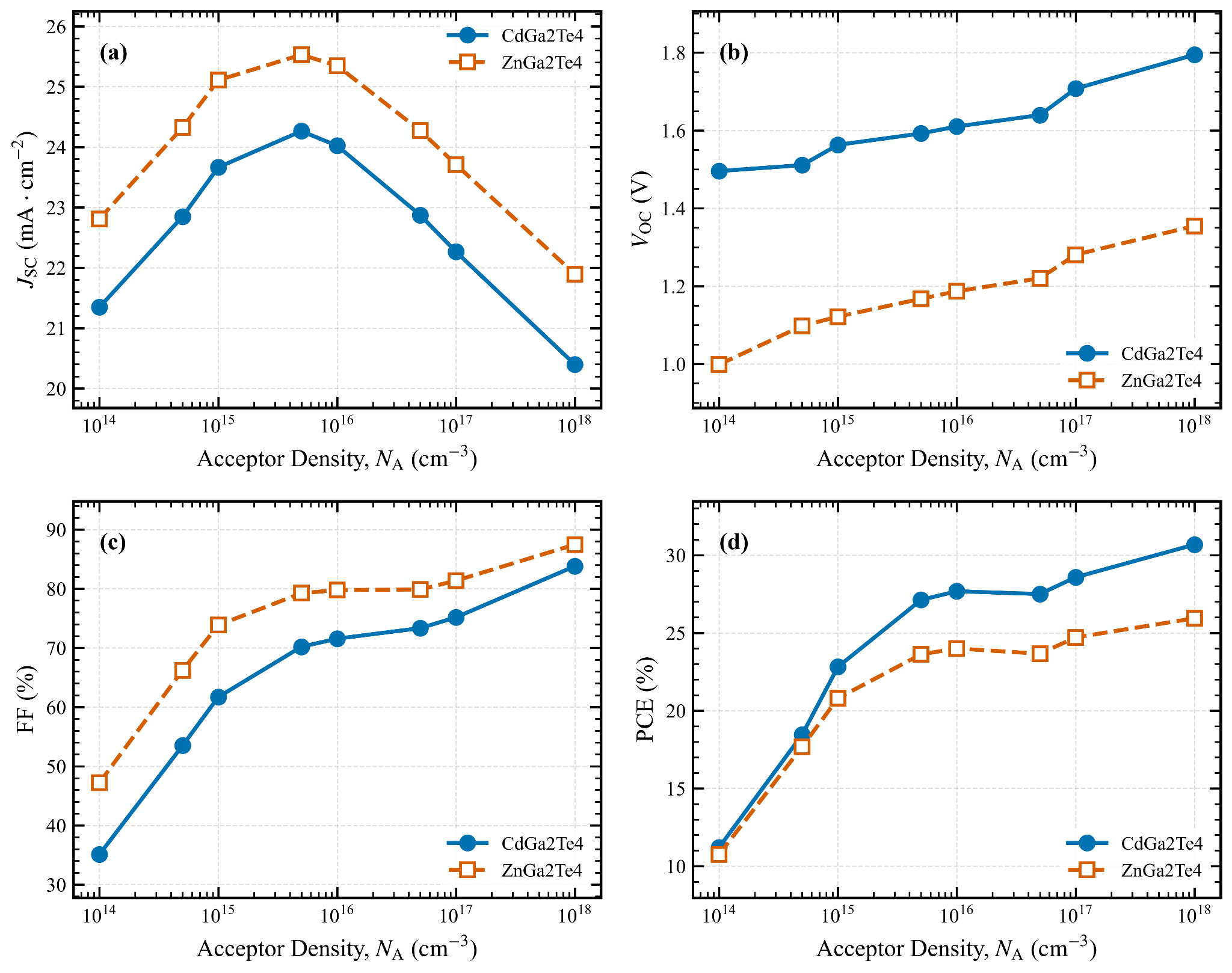}
  \caption{CHARGE-simulated dependence of (a) short-circuit current density
           ($J_{\mathrm{SC}}$), (b) open-circuit voltage ($V_{\mathrm{OC}}$), (c) fill
           factor (FF) and (d) power-conversion efficiency (PCE) on the absorber acceptor
           density ($N_{\mathrm{A}}$) for CdGa$_2$Te$_4$- and ZnGa$_2$Te$_4$-based solar
           cells. Absolute $J_{\mathrm{SC}}$ and PCE values are upper bounds
           (Section~\ref{sec:jv}).}
  \label{fig:doping}
\end{figure*}

Fig.~\ref{fig:doping} shows the photovoltaic parameters as the absorber acceptor density
$N_{\mathrm{A}}$ is increased from $10^{14}$ to $10^{18}$~cm$^{-3}$. $J_{\mathrm{SC}}$ is
non-monotonic in both devices. It rises from $\approx$21.4~mA~cm$^{-2}$ (CdGa$_2$Te$_4$)
and $\approx$22.8~mA~cm$^{-2}$ (ZnGa$_2$Te$_4$) at $10^{14}$~cm$^{-3}$ to maxima of
$\approx$24.3 and $\approx$25.5~mA~cm$^{-2}$ at $5\times10^{15}$~cm$^{-3}$, and then
decreases to $\approx$20.4 and $\approx$21.9~mA~cm$^{-2}$ at $10^{18}$~cm$^{-3}$. In
contrast, $V_{\mathrm{OC}}$ increases monotonically, from $\approx$1.50 to
$\approx$1.80~V for CdGa$_2$Te$_4$ and from $\approx$1.00 to $\approx$1.36~V for
ZnGa$_2$Te$_4$. FF rises steeply, from $\approx$35.1\% to $\approx$83.8\%
(CdGa$_2$Te$_4$) and from $\approx$47.2\% to $\approx$87.5\% (ZnGa$_2$Te$_4$), with most
of the gain occurring below $10^{15}$~cm$^{-3}$. As a result, PCE increases from
$\approx$11.2\% to $\approx$30.7\% for CdGa$_2$Te$_4$ and from $\approx$10.8\% to
$\approx$25.9\% for ZnGa$_2$Te$_4$, with a shallow plateau between $10^{16}$ and
$5\times10^{16}$~cm$^{-3}$.

These trends are governed by the width of the front space-charge region, which for a
one-sided junction scales as \cite{sze2007}
\begin{equation}
  W = \sqrt{\frac{2\varepsilon_0\varepsilon_r V_{\mathrm{bi}}}{qN_{\mathrm{A}}}} ,
  \label{eq:W}
\end{equation}
where $V_{\mathrm{bi}}$ is the built-in potential. Taking the $\approx$0.3~$\mu$m front
space-charge width in Fig.~\ref{fig:device}(c, d) at
$N_{\mathrm{A}} = 2\times10^{16}$~cm$^{-3}$ as a reference, Eq.~(\ref{eq:W}) implies a
width of $\approx$4~$\mu$m at $10^{14}$~cm$^{-3}$. The 1.5~$\mu$m absorber is then fully
depleted, and the built-in potential is spread over the entire layer. The resulting weak
field makes carrier collection strongly bias dependent, which accounts for the very low FF
at low doping. Increasing $N_{\mathrm{A}}$ up to $5\times10^{15}$~cm$^{-3}$ concentrates
the field and raises the absorber conductivity, so $J_{\mathrm{SC}}$ and FF both improve.
Beyond this point, the space-charge width keeps shrinking, by a factor of $\approx$14
between $5\times10^{15}$ and $10^{18}$~cm$^{-3}$. Because the electron diffusion length
is only $\approx$0.2~$\mu$m, the collection zone contracts and more electrons generated
deeper in the absorber recombine, so $J_{\mathrm{SC}}$ falls. Similar trade-offs between
field-assisted collection and recombination have been reported in numerical studies of
thin-film absorbers \cite{isha2023cmts,rahman2022cdte,gezgin2026}.

The rise of $V_{\mathrm{OC}}$ can be understood from the Shockley diode equation of an
illuminated cell \cite{sze2007},
\begin{equation}
  J = J_0\left[\exp\!\left(\frac{qV}{n_{\mathrm{id}}k_{\mathrm{B}}T}\right)-1\right]
      - J_{\mathrm{SC}} ,
  \label{eq:shockley}
\end{equation}
where $J_0$ is the dark saturation current density and $n_{\mathrm{id}}$ is the ideality
factor. Setting $J = 0$ gives
\begin{equation}
  V_{\mathrm{OC}} = \frac{n_{\mathrm{id}}k_{\mathrm{B}}T}{q}
  \ln\!\left(\frac{J_{\mathrm{SC}}}{J_0} + 1\right),
  \qquad J_0 \propto \frac{n_i^{2}}{N_{\mathrm{A}}} ,
  \label{eq:voc}
\end{equation}
so $V_{\mathrm{OC}}$ increases as the equilibrium minority-carrier density
$n_i^2/N_{\mathrm{A}}$ decreases. For $n_{\mathrm{id}} = 1$ at 300~K,
Eq.~(\ref{eq:voc}) predicts a gain of 59.5~mV per decade of $N_{\mathrm{A}}$. The simulated
gains are larger, $\approx$75~mV per decade for CdGa$_2$Te$_4$ and $\approx$89~mV per
decade for ZnGa$_2$Te$_4$ over the four decades studied. This excess suggests that higher
doping also suppresses recombination channels beyond the bulk term, most plausibly at the
interfaces, where stronger doping reduces the hole depletion at the CuAlO$_2$/absorber
interface (Section~\ref{sec:alignment}). Within the present model, the highest PCE occurs
at $10^{18}$~cm$^{-3}$. However, the mobilities and lifetimes in Table~\ref{tab:charge}
were held constant, whereas in real materials heavy doping generally reduces both. The
decline of $J_{\mathrm{SC}}$ above $5\times10^{15}$~cm$^{-3}$ already signals a collection
penalty, so a doping window of $10^{16}$--$10^{17}$~cm$^{-3}$ is the more robust design
target. In this window, the simulated PCE remains between $\approx$27.5\% and
$\approx$28.6\% for CdGa$_2$Te$_4$ and between $\approx$23.7\% and $\approx$24.7\% for
ZnGa$_2$Te$_4$, while $J_{\mathrm{SC}}$ stays above 22~mA~cm$^{-2}$.

\subsubsection{Optical field distribution and absorber thickness}
\label{sec:thickness}

\begin{figure*}[pos=t]
  \centering
  \includegraphics[width=\textwidth]{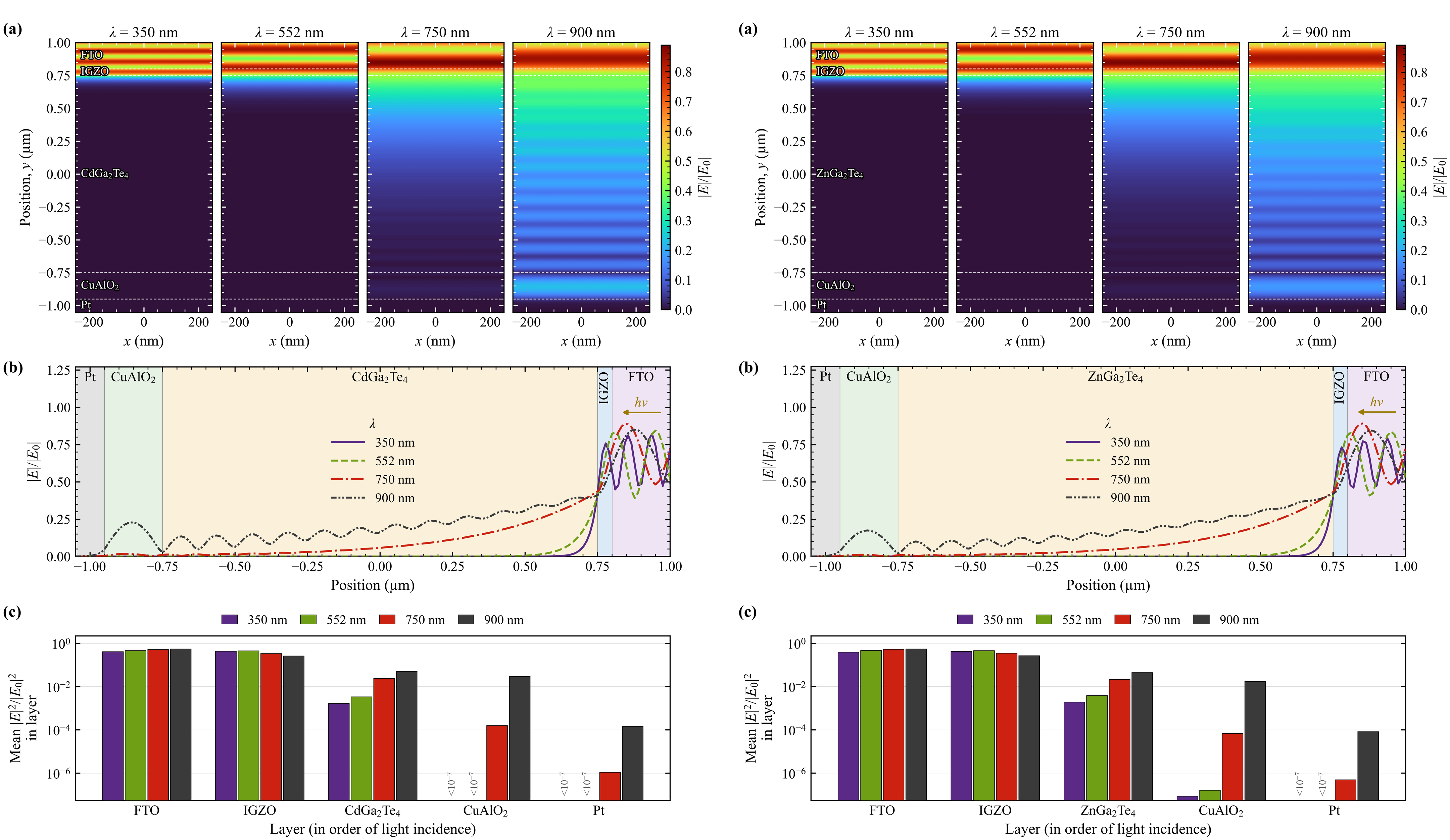}
  \caption{FDTD optical electric-field analysis of the CdGa$_2$Te$_4$-based device at
           $\lambda$ = 350, 552, 750 and 900~nm: (a) spatial maps of the normalized
           field amplitude $|E|/|E_0|$, (b) depth profiles of $|E|/|E_0|$ at $x = 0$, and
           (c) layer-averaged field intensity $|E|^2/|E_0|^2$. Corresponding results for ZnGa$_2$Te$_4$ show a qualitatively similar pattern with a peak field amplitude of $\approx$0.17 in CuAlO$_2$ at 900~nm.}
  \label{fig:efield}
\end{figure*}

The thickness dependence of the device is rooted in the optical field distribution shown
in Fig.~\ref{fig:efield} for four representative wavelengths. At 350 and 552~nm, the
normalized field $|E|/|E_0|$ oscillates between $\approx$0.4 and $\approx$0.85 within FTO
and IGZO, which is the interference pattern responsible for the spectral fringes in
Fig.~\ref{fig:ART}. Inside the absorber, the field at these wavelengths decays to near zero
within $\approx$0.1--0.15~$\mu$m, and the layer-averaged intensity $|E|^2/|E_0|^2$ in
CuAlO$_2$ and Pt falls below $\approx10^{-7}$. Short-wavelength light is therefore absorbed
entirely in the front part of the absorber, and the back layers are optically inactive at
these wavelengths. At 750~nm, the field decays gradually across the whole absorber, and
its mean intensity there is roughly an order of magnitude higher than at 350~nm. At
900~nm, the field reaches the rear of the device and forms a standing wave in the
absorber, with a peak amplitude of $\approx$0.23 inside CuAlO$_2$ in the CdGa$_2$Te$_4$
device ($\approx$0.17 for ZnGa$_2$Te$_4$). The period of a standing wave formed
against a mirror is
\begin{equation}
  \Lambda = \frac{\lambda}{2n} .
  \label{eq:standing}
\end{equation}
The fringe spacing of $\approx$0.125~$\mu$m in the depth profile at 900~nm gives
$n \approx 3.6$ at this wavelength. This value is consistent with the PBEsol refractive
index near 1.38~eV (Fig.~\ref{fig:optical2}(c)), which confirms that unshifted PBEsol
optical constants were used (Section~\ref{sec:fdtd-method}). Long-wavelength light thus
experiences the absorber as a Fabry--P\'erot cavity bounded by the front stack and the Pt
mirror \cite{pettersson1999,burkhard2010}, so its absorption depends on the absorber
thickness.

\begin{figure*}[pos=t]
  \centering
  \includegraphics[width=\textwidth]{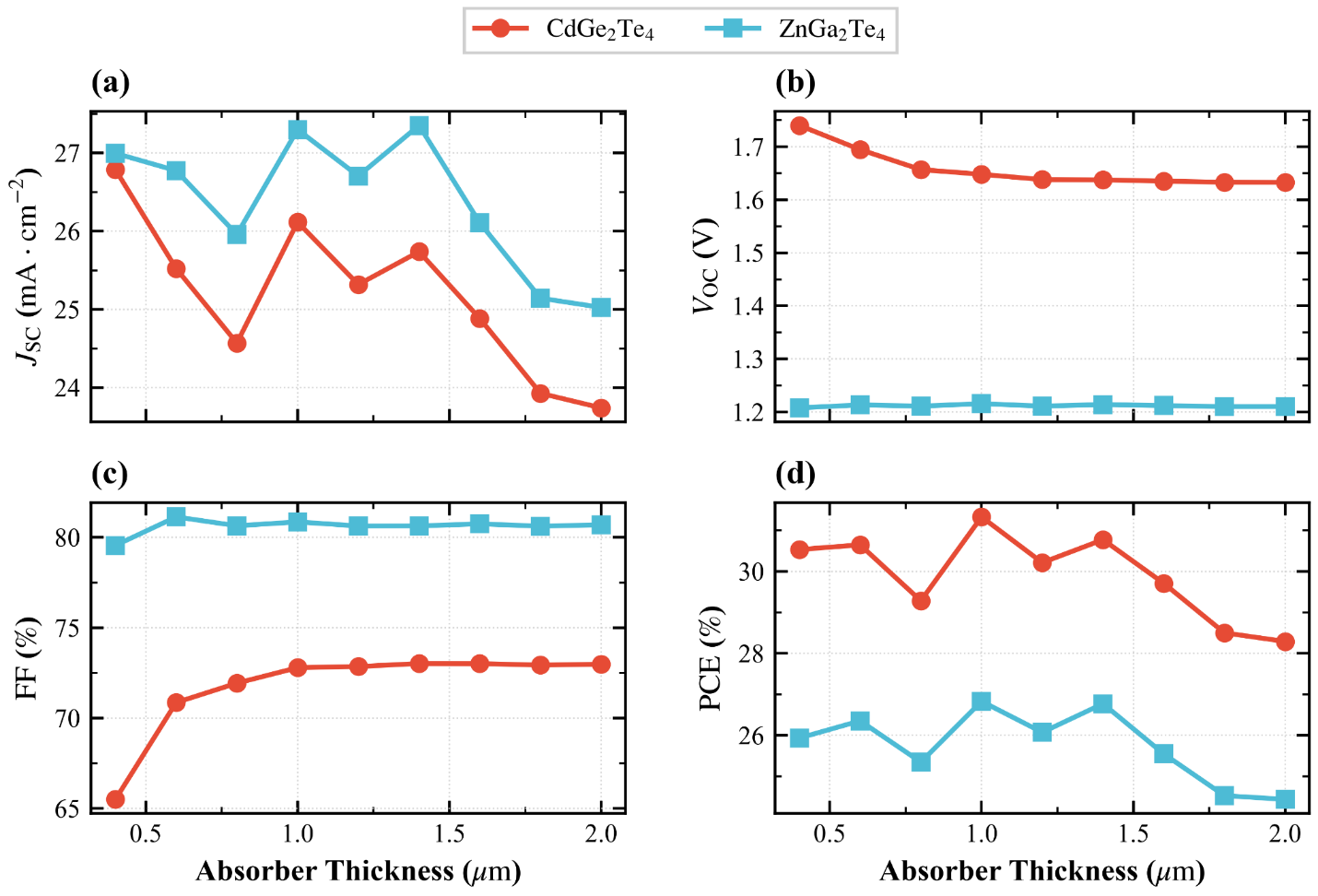}
  \caption{CHARGE-simulated effect of absorber thickness on (a) $J_{\mathrm{SC}}$,
           (b) $V_{\mathrm{OC}}$, (c) FF and (d) PCE of CdGa$_2$Te$_4$- and
           ZnGa$_2$Te$_4$-based solar cells. The FDTD generation profile was recomputed for
           each thickness. Absolute $J_{\mathrm{SC}}$ and PCE values are upper bounds
           (Section~\ref{sec:jv}).}
  \label{fig:thickness}
\end{figure*}

Fig.~\ref{fig:thickness} shows the effect of absorber thickness between 0.4 and
2.0~$\mu$m. $J_{\mathrm{SC}}$ combines a slow decline with an oscillation. For
CdGa$_2$Te$_4$, it falls from $\approx$26.8~mA~cm$^{-2}$ at 0.4~$\mu$m to
$\approx$23.7~mA~cm$^{-2}$ at 2.0~$\mu$m, with local maxima of $\approx$26.1 and
$\approx$25.7~mA~cm$^{-2}$ at 1.0 and 1.4~$\mu$m. ZnGa$_2$Te$_4$ follows the same pattern
at a higher level, with maxima of $\approx$27.3 and $\approx$27.4~mA~cm$^{-2}$ at 1.0 and
1.4~$\mu$m and $\approx$25.0~mA~cm$^{-2}$ at 2.0~$\mu$m. The oscillation arises from the
thickness-dependent standing-wave pattern of long-wavelength light described above;
because these wavelengths lie below the HSE06 gaps, it is expected to diminish with
gap-corrected optical constants. The decline has an electrical origin. Because only the
front $\approx$0.5~$\mu$m lies within the electron collection length, additional thickness
adds little collectable generation; it also moves the Pt mirror, which in thin absorbers
returns long-wavelength light into the collection zone, further from the junction.
Comparable trade-offs between absorption and collection have been reported for other
thin-film absorbers \cite{isha2023cmts,rahman2022cdte}. Thickness was sampled in
0.2~$\mu$m steps, which is coarser than the $\approx$0.125~$\mu$m fringe period derived
from Eq.~(\ref{eq:standing}), so the exact positions of the local maxima are not resolved.

The two absorbers differ markedly in their voltage response. For CdGa$_2$Te$_4$,
$V_{\mathrm{OC}}$ decreases from $\approx$1.74~V at 0.4~$\mu$m to $\approx$1.65~V at
1.0~$\mu$m and then saturates at $\approx$1.63~V. In thin absorbers, the
electron-reflecting rear barrier ($\Delta E_{\mathrm{C}} = +1.84$~eV) lies within a few
diffusion lengths of the junction and limits minority-carrier loss; once the thickness
greatly exceeds $L_n$, the rear contact no longer influences the junction and
$V_{\mathrm{OC}}$ becomes thickness-independent. For ZnGa$_2$Te$_4$, $V_{\mathrm{OC}}$
remains at $\approx$1.21~V over the entire range. This insensitivity to thickness suggests
that the voltage of the ZnGa$_2$Te$_4$ device is limited by a mechanism that does not
scale with absorber volume, such as interfacial recombination. The FF of CdGa$_2$Te$_4$
drops to $\approx$65.5\% at 0.4~$\mu$m but stabilizes at $\approx$73\% above 1.0~$\mu$m,
whereas the FF of ZnGa$_2$Te$_4$ stays between $\approx$79.5\% and $\approx$81.1\%. The
low FF of the thinnest CdGa$_2$Te$_4$ cell is consistent with the band diagram in
Fig.~\ref{fig:device}(c): the front ($\approx$0.3~$\mu$m) and rear
($\approx$0.1--0.15~$\mu$m) space-charge regions together span nearly the full
0.4~$\mu$m, leaving no quasi-neutral region, so collection becomes strongly bias
dependent. Because $V_{\mathrm{OC}}$ and FF saturate above 1.0~$\mu$m, PCE follows $J_{\mathrm{SC}}$.
Within the sampled range, the highest simulated PCE is $\approx$31.3\% for CdGa$_2$Te$_4$
and $\approx$26.8\% for ZnGa$_2$Te$_4$, both at 1.0~$\mu$m, with a second maximum at
1.4~$\mu$m ($\approx$30.8\% and $\approx$26.8\%); PCE then decreases to $\approx$28.3\%
and $\approx$24.4\% at 2.0~$\mu$m. These absolute values inherit the $J_{\mathrm{SC}}$
overestimate discussed in Section~\ref{sec:jv}, and the local maxima arise from sub-gap
interference. The robust conclusion is that absorbers much thicker than $\approx$1~$\mu$m
add little collectable generation, because electron collection is confined to
$\approx$0.5~$\mu$m from the junction. A thinner absorber also reduces the consumption of
scarce Te.

\FloatBarrier

\subsubsection{Current density--voltage and power density--voltage characteristics}
\label{sec:jv}

\begin{figure*}[pos=t]
  \centering
  \includegraphics[width=0.85\textwidth]{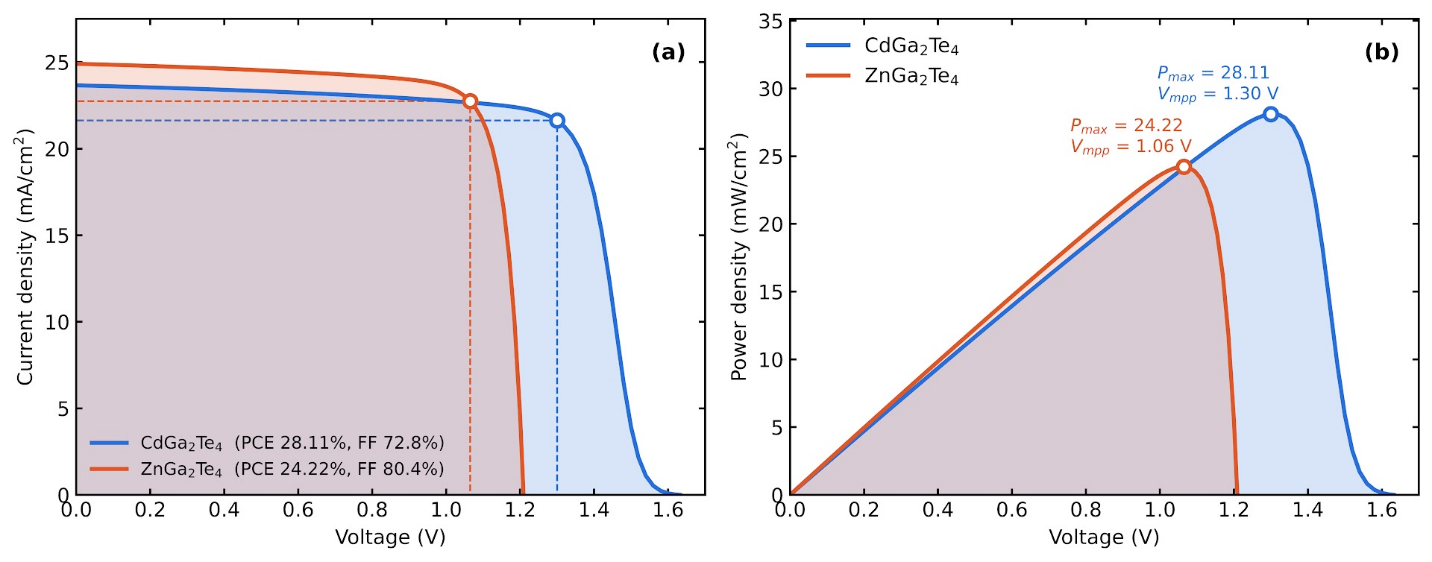}
  \caption{CHARGE-simulated (a) current density--voltage ($J$--$V$) and (b) power
           density--voltage ($P$--$V$) characteristics of the CdGa$_2$Te$_4$- and
           ZnGa$_2$Te$_4$-based solar cells under AM1.5G illumination (100~mW~cm$^{-2}$),
           computed with the FDTD generation profiles as the optical source term. The
           maximum power points are marked.}
  \label{fig:jv}
\end{figure*}

Fig.~\ref{fig:jv} presents the illuminated $J$--$V$ and $P$--$V$ characteristics of the
two devices in the baseline configuration of Table~\ref{tab:charge}. The CHARGE simulations
give $J_{\mathrm{SC}} \approx 23.7$~mA~cm$^{-2}$, $V_{\mathrm{OC}} \approx 1.63$~V,
FF~=~72.8\% and PCE~=~28.11\% for the CdGa$_2$Te$_4$ cell ($V_{\mathrm{mpp}} = 1.30$~V,
$J_{\mathrm{mpp}} \approx 21.6$~mA~cm$^{-2}$), and $J_{\mathrm{SC}} \approx
24.9$~mA~cm$^{-2}$, $V_{\mathrm{OC}} \approx 1.21$~V, FF~=~80.4\% and PCE~=~24.22\% for
the ZnGa$_2$Te$_4$ cell ($V_{\mathrm{mpp}} = 1.06$~V,
$J_{\mathrm{mpp}} \approx 22.8$~mA~cm$^{-2}$).

These values must be checked against the detailed-balance (Shockley--Queisser) limit
\cite{shockley1961}. For a single junction under AM1.5G illumination at 300~K, the
limiting $J_{\mathrm{SC}}$ is $\approx$10.4~mA~cm$^{-2}$ for $E_g = 2.21$~eV and
$\approx$19.3~mA~cm$^{-2}$ for $E_g = 1.81$~eV, and the limiting efficiencies are
$\approx$18.4\% and $\approx$26.9\%, respectively (calculated from the ASTM G173-03
spectrum \cite{astmg173}, consistent with the tabulation in Ref.~\cite{ruhle2016}). The
simulated $J_{\mathrm{SC}}$ exceeds these limits by factors of $\approx$2.3
(CdGa$_2$Te$_4$) and $\approx$1.3 (ZnGa$_2$Te$_4$), and the simulated PCE of
CdGa$_2$Te$_4$ exceeds its limiting efficiency. The excess arises because the generation
rate imported from FDTD includes photons with energies between the PBEsol and HSE06 gaps
(Sections~\ref{sec:fdtd-method} and \ref{sec:absorption}), whereas CHARGE assigns the
HSE06 gaps to the absorbers. For comparison, complete absorption of all photons in the
300--900~nm window would give 33.7~mA~cm$^{-2}$. The absolute $J_{\mathrm{SC}}$ and PCE
values in Figs.~\ref{fig:doping}, \ref{fig:thickness} and \ref{fig:jv} and
Table~\ref{tab:summary} are therefore not physically attainable for the HSE06 gaps and are reported as upper bounds. $V_{\mathrm{OC}}$ and FF, by contrast, are set mainly by the band gaps,
band offsets and recombination parameters in CHARGE and depend only weakly
($V_{\mathrm{OC}}$ logarithmically, Eq.~(\ref{eq:voc})) on $J_{\mathrm{SC}}$. Their
analysis below is therefore expected to remain qualitatively valid. CdGa$_2$Te$_4$ shows the higher simulated PCE. The origin of this advantage becomes clear
when the voltage is referenced to the band gap through the open-circuit voltage deficit,
\begin{equation}
  W_{\mathrm{OC}} = \frac{E_g}{q} - V_{\mathrm{OC}} .
  \label{eq:woc}
\end{equation}
Using the band gaps in Table~\ref{tab:charge}, $W_{\mathrm{OC}}$ is $\approx$0.58~V for
CdGa$_2$Te$_4$ and $\approx$0.60~V for ZnGa$_2$Te$_4$. The difference in $V_{\mathrm{OC}}$
($\approx$0.42~V) thus closely tracks the difference in band gap (0.40~eV), and the two
absorbers convert their band gaps into voltage with almost equal effectiveness
\cite{li2025}. The efficiency lead of CdGa$_2$Te$_4$ is a consequence of its wider gap,
not of lower recombination. The fill factor tells a different story. The ideal,
recombination-limited fill factor can be estimated from the normalized open-circuit
voltage \cite{green1981}:
\begin{equation}
  \mathrm{FF}_0 = \frac{v_{\mathrm{oc}} - \ln\left(v_{\mathrm{oc}} + 0.72\right)}
                      {v_{\mathrm{oc}} + 1} ,
  \qquad
  v_{\mathrm{oc}} = \frac{qV_{\mathrm{OC}}}{n_{\mathrm{id}}k_{\mathrm{B}}T} .
  \label{eq:ff0}
\end{equation}

\FloatBarrier

For $n_{\mathrm{id}} = 1$ at 300~K, Eq.~(\ref{eq:ff0}) gives $\mathrm{FF}_0 \approx 92.0\%$
for CdGa$_2$Te$_4$ and $\approx$89.8\% for ZnGa$_2$Te$_4$. The simulated FF values
therefore fall short of their ideal limits by $\approx$19.2 percentage points for
CdGa$_2$Te$_4$ but only $\approx$9.4 points for ZnGa$_2$Te$_4$. Separating the current and
voltage contributions locates the source of this loss. The current ratio
$J_{\mathrm{mpp}}/J_{\mathrm{SC}}$ is similar in both devices ($\approx$0.91 and
$\approx$0.92), but the voltage ratio $V_{\mathrm{mpp}}/V_{\mathrm{OC}}$ is $\approx$0.80
for CdGa$_2$Te$_4$ against $\approx$0.88 for ZnGa$_2$Te$_4$. The CdGa$_2$Te$_4$ curve
accordingly shows a soft knee as it approaches $V_{\mathrm{OC}}$ (Fig.~\ref{fig:jv}(a)),
which is characteristic of a series-resistance-like or barrier-limited extraction loss.
This observation is consistent with the hole-transport barrier and the $E_{\mathrm{Fp}}$
step at the CdGa$_2$Te$_4$/CuAlO$_2$ interface (Section~\ref{sec:alignment}), and with the
lower hole mobility of CdGa$_2$Te$_4$ (11.249 versus 17.648~cm$^2$~V$^{-1}$~s$^{-1}$). Both
curves also show a gradual decline in current of $\approx$4\% between 0 and 1.0~V,
reflecting bias-dependent collection associated with the short electron lifetime.

These results identify the rear contact as the main lever for improving CdGa$_2$Te$_4$.
As a simple estimate, raising its FF from 72.8\% to the 80.4\% obtained for
ZnGa$_2$Te$_4$, at unchanged $J_{\mathrm{SC}}$ and $V_{\mathrm{OC}}$, would increase the
PCE by $\approx$10\% in relative terms. In practice, this could be pursued by raising the
effective work function of the hole contact or by inserting an interlayer that reduces the
valence-band cliff \cite{minemoto2015}.

\subsubsection{Comparative evaluation of optical and electrical metrics}
\label{sec:comparison}

\begin{table}[!htb]
  \centering
  \caption{Optical, interfacial and electrical metrics of the CdGa$_2$Te$_4$- and
           ZnGa$_2$Te$_4$-based devices. Values marked $\approx$ were read from or derived
           from the simulated curves. $J_{\mathrm{SC}}$, $J_{\mathrm{mpp}}$ and PCE exceed
           the detailed-balance limit and are upper bounds (Section~\ref{sec:jv}).}
  \label{tab:summary}
  \footnotesize
  \setlength{\tabcolsep}{2.5pt}
  \begin{tabular}{p{3.9cm}cc}
    \toprule
    Metric & CdGa$_2$Te$_4$ & ZnGa$_2$Te$_4$ \\
    \midrule
    Band gap used in CHARGE, $E_g$ (eV)            & 2.21                & 1.81 \\
    Absorption maxima (nm)                         & 337, 431, 650       & 335, 430, 651 \\
    Minimum absorptance, 300--900~nm               & $\approx$0.64       & $\approx$0.66 \\
    $\Delta E_{\mathrm{C}}$, absorber/IGZO (eV)    & +0.15               & +0.33 \\
    $\Delta E_{\mathrm{V}}$, absorber/IGZO (eV)    & $-0.69$             & $-0.91$ \\
    $\Delta E_{\mathrm{C}}$, CuAlO$_2$/absorber (eV) & +1.84             & +2.01 \\
    $\Delta E_{\mathrm{V}}$, CuAlO$_2$/absorber (eV) & +0.59             & +0.36 \\
    Electron diffusion length, $L_n$ (nm)          & $\approx$198        & $\approx$213 \\
    Hole diffusion length, $L_p$ (nm)              & $\approx$171        & $\approx$214 \\
    $J_{\mathrm{SC}}$ (mA~cm$^{-2}$)               & $\approx$23.7       & $\approx$24.9 \\
    $V_{\mathrm{OC}}$ (V)                          & $\approx$1.63       & $\approx$1.21 \\
    FF (\%)                                        & 72.8                & 80.4 \\
    PCE (\%)                                       & 28.11               & 24.22 \\
    $V_{\mathrm{mpp}}$ (V) / $J_{\mathrm{mpp}}$ (mA~cm$^{-2}$) & 1.30 / $\approx$21.6 & 1.06 / $\approx$22.8 \\
    Voltage deficit, $W_{\mathrm{OC}}$ (V)         & $\approx$0.58       & $\approx$0.60 \\
    Ideal fill factor, FF$_0$ (\%)                 & $\approx$92.0       & $\approx$89.8 \\
    Detailed-balance PCE limit for $E_g$ (\%)      & $\approx$18.4       & $\approx$26.9 \\
    Highest PCE in $N_{\mathrm{A}}$ sweep, at $10^{18}$~cm$^{-3}$ (\%) & $\approx$30.7 & $\approx$25.9 \\
    Highest PCE in thickness sweep, at 1.0~$\mu$m (\%) & $\approx$31.3 & $\approx$26.8 \\
    \bottomrule
  \end{tabular}
\end{table}

Table~\ref{tab:summary} compiles the key optical, interfacial and electrical metrics of
the two devices. Optically, the two absorbers are nearly indistinguishable: their
absorption maxima coincide to within 2~nm, their minimum absorptance differs by less than
0.03, and both are limited by front-surface reflection rather than transmission. This
similarity arises because the spectral response is shaped mainly by the shared front stack
and Pt mirror. The differences between the devices are therefore electrical. CdGa$_2$Te$_4$
benefits from its wider band gap, which yields a $\approx$0.42~V higher $V_{\mathrm{OC}}$
at an almost identical voltage deficit. ZnGa$_2$Te$_4$ benefits from higher carrier
mobilities, longer diffusion lengths and a smaller rear valence-band cliff, which give it
a $\approx$5\% higher $J_{\mathrm{SC}}$ and a 7.6 percentage-point higher FF. In the
present simulations, the voltage advantage outweighs the current and fill-factor
penalties, so CdGa$_2$Te$_4$ shows the higher PCE. This ranking may not survive the
correction of the optical constants: the correction will reduce $J_{\mathrm{SC}}$ more
strongly for the wider-gap CdGa$_2$Te$_4$, and the detailed-balance efficiency limit is
higher for the ZnGa$_2$Te$_4$ gap ($\approx$25--27\% for 1.81--1.895~eV) than for
CdGa$_2$Te$_4$ ($\approx$18\%).

The parametric studies point to the same design priorities for both absorbers:
suppression of front-surface reflection, an absorber no thicker than $\approx$1~$\mu$m, an
acceptor density of $10^{16}$--$10^{17}$~cm$^{-3}$ and, particularly for CdGa$_2$Te$_4$,
a hole contact with better valence-band and work-function alignment. Combining
first-principles material parameters with device-level simulation in this way is
increasingly used to screen emerging absorbers before synthesis \cite{sk2025k2tl,sharma2025}.
Besides the optical-constant correction discussed above, the model neglects parasitic
series and shunt resistances and represents interface recombination only through fixed
surface recombination velocities, so the simulated efficiencies remain upper bounds for
the assumed parameter set.

\FloatBarrier
\section{Conclusions}
\label{sec:conclusions}

CdGa$_2$Te$_4$ and ZnGa$_2$Te$_4$ were investigated by first-principles calculations and
coupled FDTD--CHARGE device simulations. Both compounds are dynamically stable defect
chalcopyrites ($I\bar{4}$) with negative formation energies. They have direct gaps at
$\Gamma$ of 1.046 and 1.032~eV (PBEsol) and 2.21 and 1.895~eV (HSE06), absorption
coefficients above $10^{5}$~cm$^{-1}$ in the visible range and light electrons (0.214 and
$0.39\,m_0$). Their exciton binding energies of 18.85 and 26.81~meV are comparable to
$k_{\mathrm{B}}T$ at room temperature, so excitons dissociate readily, more easily in
CdGa$_2$Te$_4$. Constant-relaxation-time transport calculations give Seebeck coefficients
of up to $\approx$255~$\mu$V~K$^{-1}$; the electronic figure of merit is an upper bound
that excludes lattice thermal conductivity and should be re-evaluated once the plotted
thermal conductivity is confirmed. Mulliken estimates place the band edges of both
compounds across the water redox potentials, with ample driving force for hydrogen
evolution but oxygen-evolution margins of only 0.16 and 0.07~eV, so both are better suited
to hydrogen evolution than to unassisted overall water splitting.

In FTO/IGZO/XGa$_2$Te$_4$/CuAlO$_2$/Pt cells, the FDTD simulations show that the FTO/IGZO
front stack controls reflection losses through thin-film interference and that
photogeneration is concentrated near the front junction. The CHARGE simulations show a
small conduction-band spike at the front interface, electron collection confined to
$\approx$0.5~$\mu$m from the junction, and an optimal acceptor density of
$10^{16}$--$10^{17}$~cm$^{-3}$. The valence-band cliff of CuAlO$_2$ limits the fill factor
of the CdGa$_2$Te$_4$ device, which makes the hole contact the main target for
optimization. The simulated photocurrents exceed the detailed-balance limit because the
FDTD optical constants retain the PBEsol gap, so absolute device metrics must be
re-established with gap-corrected optical constants before the two absorbers can be ranked.
Overall, CdGa$_2$Te$_4$ and ZnGa$_2$Te$_4$ combine suitable band gaps, strong absorption
and weakly bound excitons, which justifies their further study as thin-film absorbers.

\section*{Acknowledgement}
The authors acknowledge support from the computational facilities at the University of
Rajshahi.

\section*{CRediT authorship contribution statement}
\textbf{Md Hasan Shahriar Rifat:} Conceptualization, Methodology, Formal analysis,
Validation, Writing -- original draft, Writing -- review \& editing.
\textbf{Tanvir Khan:} Formal analysis, Investigation, Writing -- original draft.
\textbf{Ababil Hossain Fahad:} Formal analysis, Investigation, Writing -- original draft.
\textbf{Md Arafat Hossain Shourov:} Formal analysis, Investigation, Writing -- original
draft.
\textbf{Md Sahat Bin Sayed:} Formal analysis, Investigation, Writing -- original draft.
\textbf{K.M. Mehedi Hassan:} Investigation, Writing -- original draft.
\section*{Funding}
This research did not receive any specific grant from funding agencies in the public,
commercial or not-for-profit sectors.

\section*{Disclosures}
The authors declare that they have no known competing financial interests or personal
relationships that could have influenced the work reported in this article.

\section*{Data availability statement}
The data supporting the findings of this study are not publicly available because no
suitable public repository currently exists to host these files. Reasonable requests for
access to the data will be considered by the corresponding author.

\section*{Supplemental document}
See Supplement 1 for supporting content.

\bibliographystyle{unsrt}
\bibliography{ref}

\end{document}